%% file: ms.tex
\documentclass[apj]{emulateapj}
\pdfoutput=1
\usepackage{amssymb,amsmath}
\usepackage{ifxetex,ifluatex}
\usepackage[perpage, symbol*]{footmisc}
\ifxetex
  \usepackage{fontspec,xltxtra,xunicode}
  \defaultfontfeatures{Mapping=tex-text,Scale=MatchLowercase}
\else
  \ifluatex
    \usepackage{fontspec}
    \defaultfontfeatures{Mapping=tex-text,Scale=MatchLowercase}
  \else
    \usepackage[utf8]{inputenc}
  \fi
\fi
\usepackage{graphicx}
\makeatletter
\ifxetex
  \usepackage[setpagesize=false, 
              unicode=false, 
              xetex,
              colorlinks=true,
              linkcolor=blue]{hyperref}
\else
  \usepackage[unicode=true,
              colorlinks=true,
              linkcolor=blue]{hyperref}
\fi
\hypersetup{breaklinks=true, pdfborder={0 0 0}}

\setfnsymbol{lamport}
\newcommand{\kms}[0]{km s$^{-1}$}

\newcommand{\coa}[0]{$^{12}$CO (J=1-0)}
\newcommand{\cob}[0]{$^{12}$CO (J=3-2)}
\newcommand{\coc}[0]{$^{13}$CO (J=1-0)}

\DeclareMathOperator*{\argmax}{arg\,max}

\begin{document}
\slugcomment{ApJ in Press}

\title{Quantifying Observational Projection Effects Using Molecular Cloud Simulations}
\shorttitle{Projection Effects in Molecular Clouds}
\shortauthors{Beaumont et al.}
\author{Christopher N. Beaumont$^{1,2}$, Stella S. R. Offner$^{3}$\footnote{Hubble Fellow}, Rahul Shetty$^4$, Simon C. O. Glover$^4$, Alyssa A. Goodman$^2$}
\affil{$^1$Institute for Astronomy, University of Hawai'i, 2680 Woodlawn Drive, Honolulu HI 96822;  beaumont@ifa.hawaii.edu}
\affil{$^2$Harvard-Smithsonian Center for Astrophysics, 60 Garden St., Cambridge MA 02138}
\affil{$^3$Department of Astronomy, Yale University, New Haven, CT 06511, USA}
\affil{$^4$Zentrum f\"ur Astronomie der Universit\"at Heidelberg, Institut f\"ur Theoretische Astrophysik, Albert-Ueberle-Str. 2, 69120 Heidelberg, Germany} 

\begin{abstract}

The physical properties of molecular clouds are often measured using spectral-line observations, which provide the only probes of the clouds' velocity structure.  It is hard, though, to assess whether and to what extent intensity features in position-position-velocity (PPV) space correspond to ``real'' density structures in position-position-position (PPP) space. In this paper, we create synthetic molecular cloud spectral-line maps of simulated molecular clouds, and present a new technique for measuring the reality of individual PPV structures. Using a dendrogram algorithm, we identify hierarchical structures in both PPP and PPV space. Our procedure projects density structures identified in PPP space into corresponding intensity structures in PPV space and then measures the geometric overlap of the projected structures with structures identified from the synthetic observation.  The fractional overlap between a PPP and PPV structure quantifies how well the synthetic observation recovers information about the 3D structure. Applying this machinery to a set of synthetic observations of CO isotopes, we  measure how well spectral-line measurements recover mass, size, velocity dispersion, and virial parameter for a simulated star-forming region. By disabling various steps of our analysis, we investigate how much opacity, chemistry, and gravity affect measurements of  physical properties extracted from PPV cubes.  For the simulations used here, which offer a decent, but not perfect, match to the properties of a star-forming region like Perseus, our results suggest that superposition induces a $\sim 40$\% uncertainty in masses, sizes, and velocity dispersions derived from \coc. As would be expected, superposition and confusion is worst in regions where the filling factor of emitting material is large.  The virial parameter is most affected by superposition, such that estimates of the virial parameter derived from PPV and PPP information typically disagree by a factor of $\sim 2$. This uncertainty makes it particularly difficult to judge whether gravitational or kinetic energy dominate a given region, since the majority of virial parameter measurements fall within a factor of 2 of the equipartition level $\alpha \sim 2$.

\end{abstract}


\maketitle

\section{Introduction}

All Galactic star formation occurs within molecular clouds \citep{mckee07}. Since the processes that
form and sculpt molecular clouds set the initial conditions for star
formation, the spatial and kinematic structure of molecular clouds provides clues 
about the star formation process.

CO is the most-utilized tracer of molecular cloud structure. Though molecular hydrogen
and atomic helium are 10$^{3-4}$ times more abundant than CO, neither radiates
efficiently in molecular clouds. The rotational transitions of CO, on the other hand, 
are easily excited at typical molecular cloud temperatures (10-20 K) and densities ($\sim100 ~{\rm cm}^{-3}$),
and are readily observed in the sub-mm and far infrared. $^{12}$CO is easily observed at low densities ($n \sim 100$ cm$^{-3}$), but is often optically thick; $^{13}$CO is $\sim$ 70 times less abundant than $^{12}$CO, and remains optically thin to higher volume density substructures \citep{http://adsabs.harvard.edu/abs/2010MNRAS.405..759D, http://adsabs.harvard.edu/abs/1999RPPh...62..143W, http://adsabs.harvard.edu/abs/1982ApJ...262..590F}. $^{13}$CO emission is associated observationally with gas at $n\gtrsim 10^3$ cm$^{-3}$.

Ideally, the full six-dimensional spatial-kinematic information would be available for studying molecular cloud structure.  Unfortunately, observations can only provide either two-dimensional information of the intensity in the plane of the sky or three-dimensional intensity information as a function of 2D space and line-of-sight velocity. For accurately interpreting observations, therefore, it is necessary to thoroughly understand the translation of physical properties in six-dimensional space to the observed emission in position-position-velocity space.

In most analyses, researchers assume (implicitly or explicitly) that \textit{intensity features} in position-position-velocity (PPV) datasets 
correspond more or less cleanly to 3D (positition-position-position, or PPP) \textit{density structures} in a cloud (see Table \ref{tab:terms} for terminology). A typical molecular cloud analysis decomposes clouds into
one or more structures based solely on the morphology of emission in PPV space and measures the properties
of these structures. For example, this is the
analysis strategy used to measure the size dependence of velocity dispersion, mass, and virial parameter \citep{http://adsabs.harvard.edu/abs/1981MNRAS.194..809L, http://adsabs.harvard.edu/abs/1987ApJ...319..730S, http://adsabs.harvard.edu/abs/2008ApJ...686..948B}. 

Molecular cloud motions are dominated by turbulence at scales above $\sim 0.1$ pc, and have complex velocity fields. Likewise, the temperature, excitation, and abundance conditions vary throughout clouds by factors of several \citep{http://adsabs.harvard.edu/abs/2008ApJ...679..481P,  http://adsabs.harvard.edu/abs/2013ApJ...766L..17S} . All of these factors affect the morphology of emission in PPV space and present a substantial obstacle to further data analysis. This raises the question: how well do features in observational data relate to intrinsic structures in the three-dimensional cloud?

\input{terms_table.tex}

The aim of this paper is to measure how well intensity structures extracted from PPV cubes correspond to density structures in PPP space. Since we can never measure how well PPV and PPP structures match up in the observed Universe, we need to make these measurements using simulations, where complete information is available in both ``spaces'' \citep{http://adsabs.harvard.edu/abs/2011IAUS..270..511G}. 

We begin with a discussion of how observational effects can distort measurements of cloud properties in Section \ref{sec:overview}. In Section \ref{sec:method}, we describe a technique to quantify how well an observed intensity feature corresponds to a real PPP structure, by measuring the partial overlap of density structures and observational features in PPV space. This technique leverages the dendrogram algorithm to decompose hierarchical cloud structure, by tracking how iso-intensity contour surfaces nest inside one another \citep{http://adsabs.harvard.edu/abs/2008ApJ...679.1338R}. By performing radiative transfer calculations on two numerical hydrodynamic models, we construct synthetic $^{12}$CO and $^{13}$CO observations of molecular clouds (Section \ref{sec:sim}). We use the COMPLETE observations of Perseus as a point of comparison throughout this work \citep{http://adsabs.harvard.edu/abs/2006AJ....131.2921R} and compare these simulations to Perseus in detail in Section \ref{sec:perseus}.
We apply our analysis in Section \ref{sec:result} to study how well measurements of intrinsic cloud properties --  mass, size, velocity dispersion, and the virial parameter -- can be recovered from observations. 
 
\subsection{Overview of Observational Effects}
\label{sec:overview}

We begin with 
a broad overview of how cloud information can be distorted during the observation process.
A spectral line observation of a molecular cloud can be thought of as a
transformation from a set of intrinsic quantities  -- density, velocity,
temperature, chemical abundance --  to a map of intensity in PPV.
Information about the original cloud structure is lost during several
steps of this transformation. The first problem is the projection from
PPP space to the PPV space
of the observation. This step is described by the following equation:

\begin{equation}
\rho_{\rm PPV}(x, y, v) = \sum_{v_z(x, y, z) = v} \rho_{\rm PPP} (x, y, z) \left| \frac{\partial z}{\partial v_z(x, y, z)} \right| 
\label{eq:projection}
\end{equation}

where $\rho_{\rm PPV}$ is the density of material in PPV space (g cm$^{-2}$ km$^{-1}$ s), $\rho_{\rm PPP}$ is the density in PPP (g cm$^{-3}$), 
and the derivative is the standard Jacobian used when transforming densities between coordinate systems. 

\begin{figure}[h!]
\includegraphics[width=3.5in]{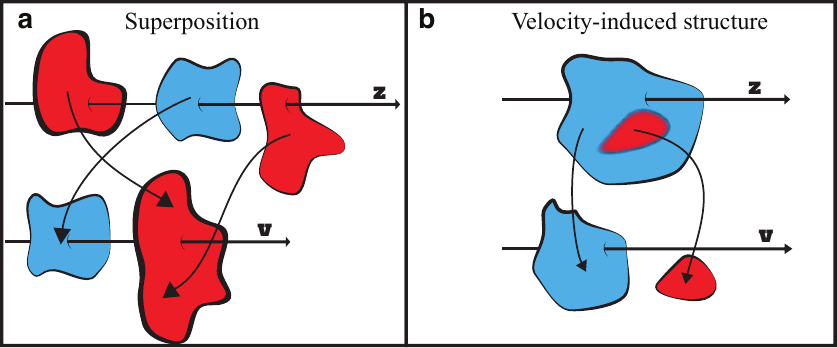}
\caption{Schematic representation of superposition and velocity-induced structures. Colors indicate velocity. Left: Three PPP structures (top) merge into 2 PPV structures (bottom), due to the similar velocity of the front and back structures. Right: A single density structure with internal velocity gradients (top) splits into two PPV structures (bottom).}
\label{fig:schematic}
\end{figure}

From the perspective of feature identification, two aspects of this transformation
break the correspondence between
PPP structures and PPV features. First, distinct positions along the
same line of sight that move at similar velocities will project to the
same region in PPV. Thus, a feature in PPV may sample
two or more density structures. This is the problem of \textit{superposition}, and is illustrated in Figure \ref{fig:schematic}a.
Second, spatial variations in $v_z$ affect
the gradient term in Equation \ref{eq:projection}, and can modulate the $\rho_{\rm PPV}$ field independently of $\rho_{\rm PPP}$. 
In other words, a single density structure can map to multiple \textit{velocity-induced} PPV features. This is shown schematically
in Figure \ref{fig:schematic}b.

In addition to projection, observations are also subject to chemical and radiative transfer
effects, which further distort the intensity field from the density
field via spatially variable excitation, abundance, ionization, and
opacity conditions \citep{http://adsabs.harvard.edu/abs/2006MNRAS.371.1865B, Lee13, http://adsabs.harvard.edu/abs/2008ApJ...679..481P}. The $\rho_{\rm PPV}$ field determines the column density and, along with the temperature, the collision rate of the gas. Both of these affect the intensity field -- the column density sets the opacity and can obscure background features, while the collision rate affects the excitation state and emissivity of the gas. Finally, observations are
subject to noise and spatial filtering, which
further degrade the data, increasing opportunities for confusion.

The net effect of these phenomena is too complicated to characterize
analytically. Instead, we turn to numerical simulations, where cloud
properties can be compared before and after synthetic ``observations''. With
simulations we also have the freedom to disable individual aspects of the
observation process, to better isolate the influence of each factor.

\subsection{Previous Work}

Several authors have investigated the relationship between PPP structures and PPV structures, using different simulations and analysis techniques. An early study by
Adler et al. (\citeyear{http://adsabs.harvard.edu/abs/1992ApJ...384...95A}) investigated structures identified in longitude-velocity diagrams of a synthetic model of the Galaxy. They found that many of these identifications were superpositions of separate PPP regions. On a smaller scale, \cite{http://adsabs.harvard.edu/abs/2002ApJ...570..734B} simulated observations of molecular cloud clumps in local thermodynamic equilibrium. They measured a number of canonical relationships 
in both PPP and PPV, including the clump mass spectrum, size-linewidth relationship, and size-density relationship. They found that the
amount of confusion due to superposition depends on both the strength and spatial scale of turbulent driving -- if turbulence induces stronger or smaller-scale density perturbations, confusion worsens. Despite confusion problems, both of these papers recovered mass-size and size-linewidth scaling relationships similar to real clouds. \cite{http://adsabs.harvard.edu/abs/1990ApJ...352..132I} also considered how cloud superposition affects the size-linewidth relationship in Galactic CO surveys, and demonstrated that the slope of the relationship is quite robust to crowding.

\cite{http://adsabs.harvard.edu/abs/2003ApJ...592..203G} analyzed the aspect ratios of molecular cloud clumps projected onto the sky. The observed distribution of cloud aspect ratios is affected
both by projection into 2D, as well as superposition of distinct cloud features. \cite{http://adsabs.harvard.edu/abs/2009ApJ...693..914O} also studied the intrinsic and projected distribution of simulated core shapes, noting that the intrinsic triaxial shape of cores is lost during projection.
Using a similar analysis,  \cite{http://adsabs.harvard.edu/abs/2002ApJ...569..280J} attempted to invert the
distribution of apparent axis ratios to recover the intrinsic shape distribution of clouds. This inversion assumed that observed cloud shapes are unaffected by superposition, however.

\cite{http://adsabs.harvard.edu/abs/2000ApJ...532..353P}  correlated PPV structures in MHD simulations with the original density field and velocity field. The features in their PPV maps resemble the patterns in the velocity field more  than the density field. There were also more small-scale PPV structures than there were PPP structures. The authors attributed these small PPV structures to the velocity-induced structures shown in Figure \ref{fig:schematic}b.

\cite{http://adsabs.harvard.edu/abs/2001ApJ...546..980O} cataloged ``observed'' structures in their 3D MHD simulations by identifying regions of contrast in 2D projections. They, too, noted that features identified in this way often consist of several superposed density structures. Even though their feature extraction process ignored line-of-sight velocity information, they reported that velocity information is often unable to disambiguate superpositions in 2D.

In a series of papers \citep{http://adsabs.harvard.edu/abs/2000ApJ...537..720L, http://adsabs.harvard.edu/abs/2004ApJ...616..943L,
http://adsabs.harvard.edu/abs/2006ApJ...652.1348L}, Lazarian et al. developed a mathematical formalism to describe how to recover statistical properties of turbulence (namely, the turbulent velocity and density power spectrum) from observations. This approach differs from the previous references in that it does not focus on the reality of observed structures. Instead,
it considers the structure functions of spectra and slices or slabs of PPV cubes. These papers derive the expected shape of observable spatial and spectral structure functions, for idealized turbulence. The advantage of this analysis is that it explicitly treats PPV superposition, though other factors like spatially varying excitation and abundance conditions are not treated.

An approach based on Principal Components Analysis has similarly been used to measure cloud statistics without identifying specific structures with clear boundaries \citep{http://adsabs.harvard.edu/abs/1997ApJ...475..173H, http://adsabs.harvard.edu/abs/2002ApJ...566..276B, http://adsabs.harvard.edu/abs/2013MNRAS.433..117B}. This method decomposes PPV datacubes into linear superpositions of ``eigenimages'' with different spatial and spectral extents. These extents, derived from the spatial and spectral autocorrelation of the decomposed data, are used to reconstruct scaling relationships like the velocity power spectrum.

Recently, \cite{http://adsabs.harvard.edu/abs/2010ApJ...712.1049S} carried out an analysis similar to the work by Ballesteros-Paredes and Mac Low (2002), and measured how projection affects the measurement of size-linewidth relationships in molecular cloud substructures. They identified structures using the dendrogram algorithm, which is explicitly designed to characterize hierarchical structures like molecular clouds. They concluded that superposition can change the power law scaling coefficient for the mass-size and size-virial relationships by $\sim \pm 0.5$, and the linewidth-size relationship by $\sim \pm 0.05$ (see Table 1 of that paper).

In summary, a broad range of work using a variety of numerical simulations has found that projection effects impact the study of cloud structures. Even though we can't explicitly measure this effect in real clouds, this work suggests it is important to quantify projection effects in the context of typical observed quantities. The analysis presented in this paper builds upon these previous studies in a few key aspects. First, we develop a method to systematically cross-match observed and real cloud structures,
for all structures in a cloud simulation. This provides a more detailed view into how structure analysis is affected by factors like superposition. We also carry out a more detailed, non-LTE radiative
transfer to better model the important effects of excitation and opacity.

\section{Methodology}
\label{sec:method}
While the net effects of projection, chemistry, radiative transfer, noise, and resolution are very difficult to study analytically, the effects can be measured empirically in simulations.  Here we describe our approach.

Consider a particular density structure, denoted by $R_i$. The structures in this work have clearly-defined boundaries (Table \ref{tab:terms}), so $R_i$ is described by a collection of voxels (3D pixels). Likewise, let $O_j$ denote the set of PPV voxels describing a particular observed feature. Using Equation \ref{eq:projection}, we can compute $\rho_{\rm PPV}(R_i)$, the distribution of $R_i$ in PPV ignoring the rest of the cloud. We can also measure $I(O_j)$, the intensity distribution of $O_j$ in PPV. We then define the similarity between PPP structure $i$ and PPV feature $j$ as

\begin{equation}
S_{ij} = \frac{\sum \rho_{\rm PPV}(R_i) \times I(O_j)}{\left[\sum \rho_{\rm PPV}(R_i)^2 \times \sum I(O_j)^2\right]^{1/2}}
\label{eq:sim}
\end{equation}
The summation is over all voxels in PPV\footnote{Under the interpretation that $\rho_{\rm PPV}(R_i)$ and $I(O_j)$ are vectors, $S_{ij}$ is their normalized dot product.}. Conceptually, $S_{ij}$ measures how much $R_i$ and $O_j$ overlap in PPV. The metric varies from 0 to 1 (0 indicating no overlap, and 1 indicating complete overlap).  In other words, large values of $S_{ij}$ suggest that $O_j$ is the observational counterpart of $R_i$. Thus, we can match an observed feature to its likely counterpart in the density field via  
 \begin{eqnarray}
  \label{eq:match}
 M_j &=& \argmax_{i}~ S_{ij}   \\ 
 q_j &=& \max_{i} ~S_{ij}  
 \label{eq:quality}
 \end{eqnarray}  

where $M_j$ is the best-matching density counterpart for PPV structure $O_j$; it is the PPP structure $i$ which maximizes $S_{ij}$. The quality factor $q_j$ characterizes the quality of the match. When $q_j$ is small, $O_j$ has no correspondence to any density structure, and is an artifact. 

\begin{figure}
\centering
\includegraphics[width=3.2in]{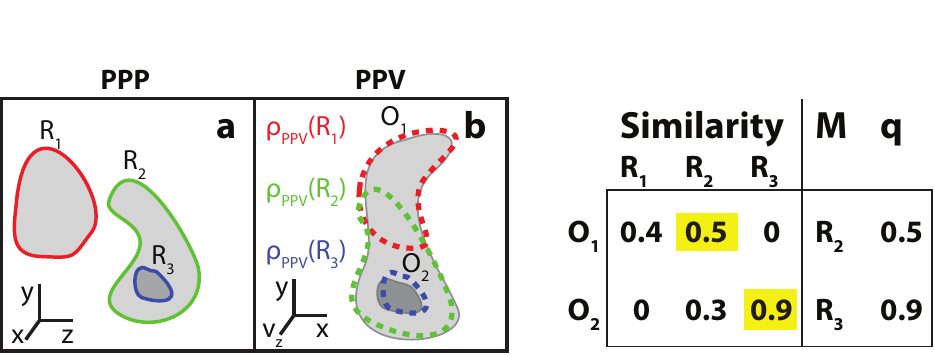}
\caption{Schematic representation of matching structures in PPV and PPP.}
\label{fig:match_schematic}
\end{figure}

Figure \ref{fig:match_schematic} depicts this process schematically, for a region with 3 PPP structures. These structures (panel a) superpose onto two PPV structures (panel b); the projection of each individual PPP structure into PPV is shown as a dotted line in panel b. The chart on the right shows the similarity matrix, as well as the match and quality for each observed structure. Structure $O_2$ matches to $R_3$ with high quality $q=0.9$. Structure $O_1$, on the other hand, is a superposition of $R_1$ and $R_2$. It matches $R_2$ slightly better than $R_1$, but the corresponding quality is low: $q=0.5$.

Equation \ref{eq:projection} describes the projection of density from PPP to PPV. However, there are two subtleties that must be addressed when carrying out this projection. The first is that the simulations in this paper are discretely sampled on a grid. In general, two neighboring voxels along a line of sight can have velocity differences greater than the velocity sampling in PPV. If each PPP cell is assigned to the single nearest velocity bin, this leads to discretization artifacts where emission ``skips over'' some velocity channels. This is described in detail in Appendix B of \cite{http://adsabs.harvard.edu/abs/2011MNRAS.415.3253S}. We circumvent this by interpolating the density field as needed, so that the velocity jump between interpolated points is always one velocity channel.

Second, the simulations assume that the velocity is constant within a cell, when in fact there should be a range of velocities at that size scale. This stems both from the thermal motion of atoms, as well as microturbulence (the turbulence at spatial scales smaller than those resolved by the simulation). Thus, each PPP location in the simulation contains material at a variety of velocities. We account for this by convolving the $\rho_{\rm PPV}$ along the velocity dimension with a Gaussian of variance $\sigma^2 = \sigma_{\rm thermal}^2 + \sigma_{\rm micro}^2$, where $\sigma_{\rm thermal}$ is the thermal linewidth and $\sigma_{\rm micro}$ is the microturbulence listed in Table \ref{tab:sim_params}.

Equations \ref{eq:match} and \ref{eq:quality} suggest a strategy for investigating projection and other observational effects in detail. Given a simulation and a hypothetical observation (described in Section \ref{sec:sim}), we catalog both the PPP and PPV structures (Section \ref{sec:structure_identification}). Then, we find $M$ and compute $q$ for all PPV structures in the simulation. These quantities allow us to investigate how well structures (and measurements of their properties) are recovered in these synthetic observations (Section \ref{sec:result}). To the extent that any simulation resembles a real cloud (Section \ref{sec:perseus} and Appendix \ref{sec:diagnostics}), this analysis offers a way to quantify otherwise un-measurable observational effects in real data. In other words, for physical conditions represented by a simulation, we can use this machinery to quantify how well mapping out any particular set of spectral line in PPV lets us estimate basic cloud properties like mass, size, line width, and virial parameter.

\subsection{Data Preparation}
\label{sec:sim}

We have applied the similarity analysis described above to two cloud simulations. Each of these simulations is meant to broadly represent the
conditions in a molecular cloud like Perseus \citep{http://adsabs.harvard.edu/abs/2006AJ....131.2921R, http://adsabs.harvard.edu/abs/2008hsf1.book..308B}. However, the mean simulation temperature, density, and line-of-sight dimension may differ from the true values in Perseus by factors of two.

The first simulation, henceforth O1, is performed with the {\sc orion} adaptive mesh refinement
(AMR) code \citep{http://adsabs.harvard.edu/abs/1998ApJ...495..821T, http://adsabs.harvard.edu/abs/1999JCoAM.109..123K}. The simulation assumes a simple isothermal equation
of state, which means that it is scale-free for density and temperature (e.g., \citealt{http://adsabs.harvard.edu/abs/2008ApJ...686.1174O}). The simulation is produced following the same procedure in
\citet{Offner13}, which we briefly summarize below. 

The simulation domain begins with a uniform density, which we perturb with
a random velocity field for two crossing times. The input field has a
flat power spectrum for large wavenumbers, $1<k<2$, and we normalize the perturbations to maintain a constant
three-dimensional Mach number, $\mathcal{M}=22$. This Mach number was chosen to reproduce the observed velocity dispersion in Perseus. After the gas achieves a well-mixed turbulent state, we turn on
self-gravity and allow collapse to proceed. The simulation has a 256$^3$
base grid, four levels of AMR refinement, and employs periodic boundary conditions. New grids are added automatically to satisfy the Jeans
criterion for a Jeans number (the ratio of cell size to the local jeans length) of 0.25 \citep{http://adsabs.harvard.edu/abs/1997ApJ...489L.179T}. When the
Jeans criterion is violated on level four within a collapsing region, a sink particle is introduced \citep{http://adsabs.harvard.edu/abs/2004ApJ...611..399K}.  Our similarity analysis is performed at half a freefall time, when $\sim 700 M_\odot$ is contained in sink particles (2.3\% of the gas).

The second simulation (hereafter S11) is an updated version of a model
 originally presented in \cite{http://adsabs.harvard.edu/abs/2011MNRAS.415.3253S} (specifically, the n100 simulation
 in that paper). It was generated using a modified version of the {\sc zeus-mp}
 MHD code \citep{http://adsabs.harvard.edu/abs/1992ApJS...80..753S,  http://adsabs.harvard.edu/abs/1992ApJS...80..791S, http://adsabs.harvard.edu/abs/2000RMxAC...9...66N, http://adsabs.harvard.edu/abs/2006ApJS..165..188H}.
 S11 differs from O1 in that it ignores gravity, includes a 5.85 $\mu$G
 magnetic field and includes treatments of the non-equilibrium heating and
 cooling of the gas, the penetration of UV radiation into the cloud, and also a
 simplified treatment of the formation and destruction of H$_2$ and CO. The original
 \cite{http://adsabs.harvard.edu/abs/2011MNRAS.415.3253S} simulation used the chemical model presented in
 \cite{http://adsabs.harvard.edu/abs/2010MNRAS.404....2G}, but the updated version presented here uses instead
 a treatment based on \cite{http://adsabs.harvard.edu/abs/1999ApJ...524..923N}, as described in \cite{http://adsabs.harvard.edu/abs/2012MNRAS.421..116G}. However, as explored in some detail in \cite{http://adsabs.harvard.edu/abs/2012MNRAS.421..116G},
 this change in chemical networks does not significantly affect the CO
 distribution in the gas. Our updated version of the \cite{http://adsabs.harvard.edu/abs/2011MNRAS.415.3253S}
 simulation also includes a number of improvements in the way in which the
 thermal evolution of the gas is modeled, as described in Appendix A of
 \cite{http://adsabs.harvard.edu/abs/2012MNRAS.421..116G}.
 
 The S11 simulation begins with uniform density which is perturbed with
 a random turbulent velocity field for three turbulent crossing times. The
 input field is similar to that in the O1 simulation, and is normalized to
 maintain a constant 3D rms velocity dispersion of $5 \: {\rm km \: s^{-1}}$.
 Converting this value to a Mach number is complicated by the fact that the
 gas in the S11 simulation is not isothermal and hence has a spatially varying
 sound speed. The volume-weighted mean Mach number is relatively low,
 ${\cal M} \simeq 6$, because much of the cloud volume is filled by warm,
 CO-poor gas with $T \sim 60$--70~K. If, however, we compute ${\cal M}$
 only for gas with more than 10\% of its carbon in CO, we find a much higher
 value, ${\cal M} \simeq 14$, as this gas is much colder, with
 $T \sim 10$--20~K.

Table \ref{tab:sim_params} summarizes the properties of each simulation.

\input{sim_params_table.tex}

We used the radiative transfer program RADMC-3D
\citep{http://adsabs.harvard.edu/abs/2012ascl.soft02015D} to generate
synthetic observations of each simulation in \coa, \cob, and \coc, using the large-velocity-gradient (LVG) approximation \citep{http://adsabs.harvard.edu/abs/1957SvA.....1..678S, http://adsabs.harvard.edu/abs/2011MNRAS.412.1686S}. The observations were
gridded to a spatial resolution of 0.1 pc pixel$^{-1}$, and velocity
resolution of 0.05 \kms. Finally, we added noise to each cube (0.6K,
0.15K and 0.25K for the \coa, \cob, and \coc\, transitions, respectively). These are
representative of the noise values of present-day cloud surveys in these
transitions
\citep{http://adsabs.harvard.edu/abs/2006AJ....131.2921R, http://adsabs.harvard.edu/abs/2010MNRAS.405..759D}.

\subsection{ Structure Identification}
\label{sec:structure_identification}

\begin{figure}[htbp]
\centering
\includegraphics[width=3.4in, trim=.9in .5in .4in 0, clip]{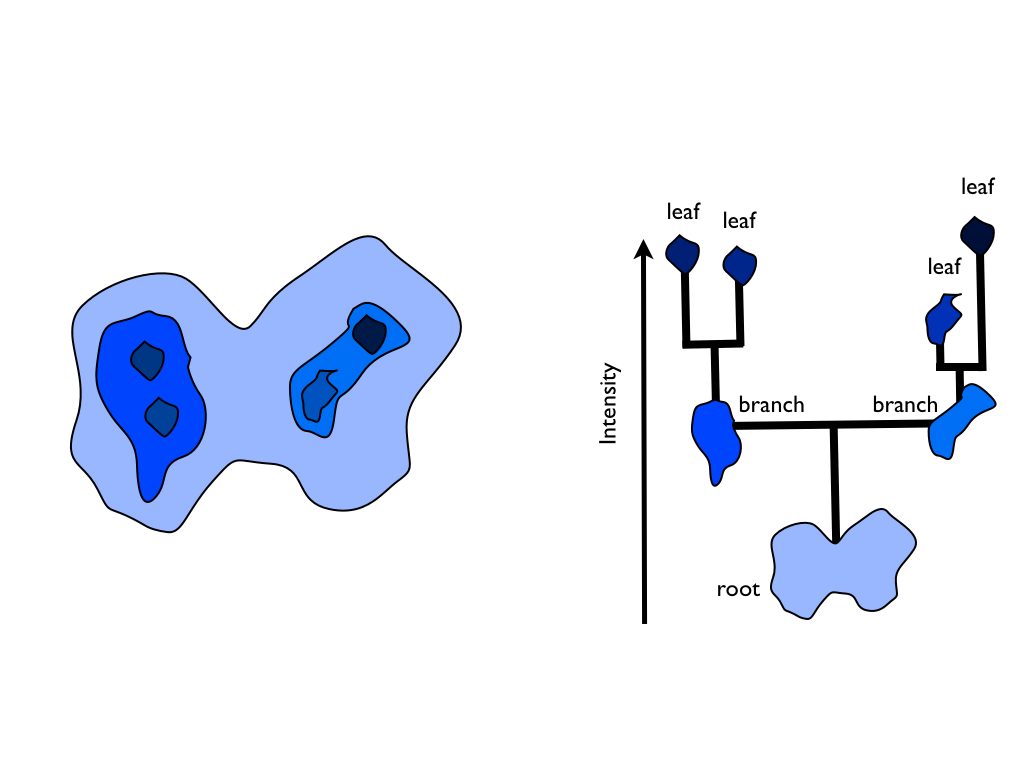}
\caption{Schematic representation of a 2D cloud (left) and its dendrogram decomposition (right). Each dendrogram structure is a closed contour in the image. The extension to 3D data is straightforward, but each structure corresponds to an iso-surface instead of a contour line.}
\label{fig:dendro_schematic}
\end{figure}

We used the dendrogram algorithm to
catalog intensity structures in each synthetic observation, as well as the density
structures in the PPP density fields. The dendrogram algorithm is
described in detail in
\cite{http://adsabs.harvard.edu/abs/2008ApJ...679.1338R}. Briefly, each
structure in a dendrogram corresponds to a surface of constant intensity
(in the observation) or density (in the PPP cube). The name dendrogram
refers to the fact that these surfaces are hierarchically nested inside
each other and representable via tree diagrams (Figure \ref{fig:dendro_schematic}). Dendrograms
capture the hierarchical structure of molecular clouds -- a clear advantage over non-hierarchical
clump-finding algorithms -- and dendrogram decompositions are
not sensitively dependent on tuning parameters of the algorithm
\citep{http://adsabs.harvard.edu/abs/2009ApJ...699L.134P}.

When constructing a dendrogram, the main freedom one has is the degree to
which structures are further decomposed into nested substructures. This
process is called ``pruning'' the dendrogram, since it amounts to
controlling how many ``branches'' (structures) are in the decomposition.
The main purpose of pruning is to suppress the extraction of
insignificant structures that are poorly resolved or are possible noise
fluctuations. A dendrogram constructed with no pruning assigns
every local intensity or density maximum (including every noise spike)
to a unique structure. The effect that pruning has on the statistical properties of a dendrogram
has been studied in detail by \cite{http://adsabs.harvard.edu/abs/2013ApJ...770..141B}.

Each dendrogram in this work is pruned such that every leaf contains a local intensity maximum that is brighter than the neighboring 7 voxels in any direction.
Each leaf (the brightest, most-compact structure in a hierarchy) also contains at least 800 voxels (for spectral line cubes in PPV) or 100 voxels (for density cubes in PPP) and contains a voxel that is at least 7$\sigma$ brighter than the contour at which the leaf merges with its neighboring structure. The PPP dendrogram is pruned less heavily than the PPV dendrograms, yielding a catalog with more structures. This prevents PPV structures from being poorly matched to PPP structures simply because the density structure decomposition is too coarsely grained. One of the convenient aspects of dendrograms is that the boundaries of non-pruned structures are independent of the pruning; in other words, while pruning can add or remove structures from a catalog, it does not affect how the included structures are defined. Thus our choice of pruning has little effect on subsequent analysis, other than to exclude from consideration the smallest cloud substructures. We explore how sensitive our analysis is to our pruning choice in Section \ref{sec:pruning_test}.

\subsection{Comparison to Perseus}
\label{sec:perseus}

\begin{figure*}[htbp]
\centering
\includegraphics[width=2.1in]{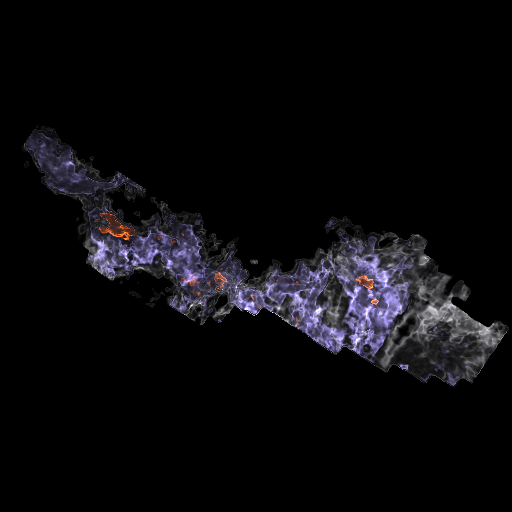}
\includegraphics[width=2.1in]{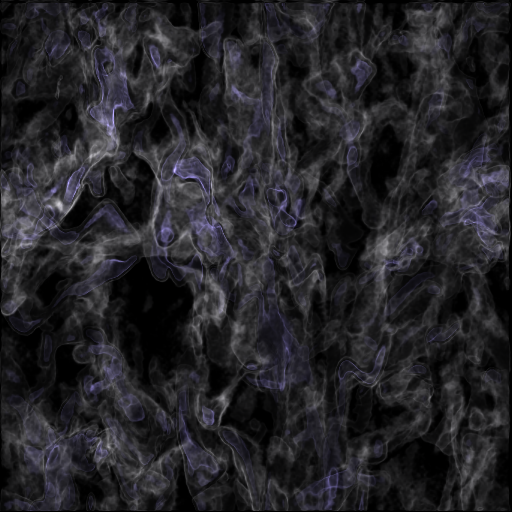}
\includegraphics[width=2.1in]{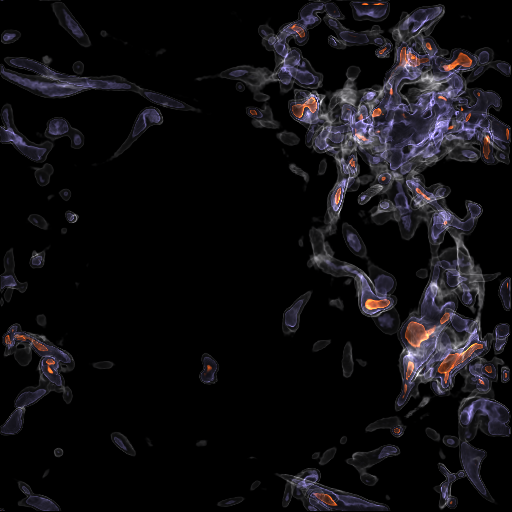}
\caption{Isosurface renderings of \coa\, emission from Perseus, O1 and S11. Isosurfaces are drawn at 3 (white), 8 (purple), and 15K (orange).}
\label{fig:volume}
\end{figure*}

We compare synthetic CO observations of the simulations to the COMPLETE Perseus data \citep{http://adsabs.harvard.edu/abs/2006AJ....131.2921R} using several
diagnostics: namely, the distribution of column density, velocity dispersion, and line intensity in various CO transitions. 
Both the O1 and S11 simulations represent the general physical properties of Perseus. Appendix A provides details about how each
quantity was extracted from the data. Neither simulation
agrees with Perseus when these diagnostics are examined in detail, but they are the closest available approximations. We discuss the limitations imposed by the suitability of the simulations in Section \ref{sec:generalize}.

Figure \ref{fig:volume} shows isosurface renderings for Perseus, O1, and S11, in the \coa\, transition. Isosurfaces are drawn at 3, 8, and 15K. The O1 simulation (panel b) stands out from the other two panels in this Figure. Compared to Perseus and S11, it has more space-filling emission at 3K and a lack of emission at 15K. 

\begin{figure*}[htbp]
\centering
\includegraphics[width=7in]{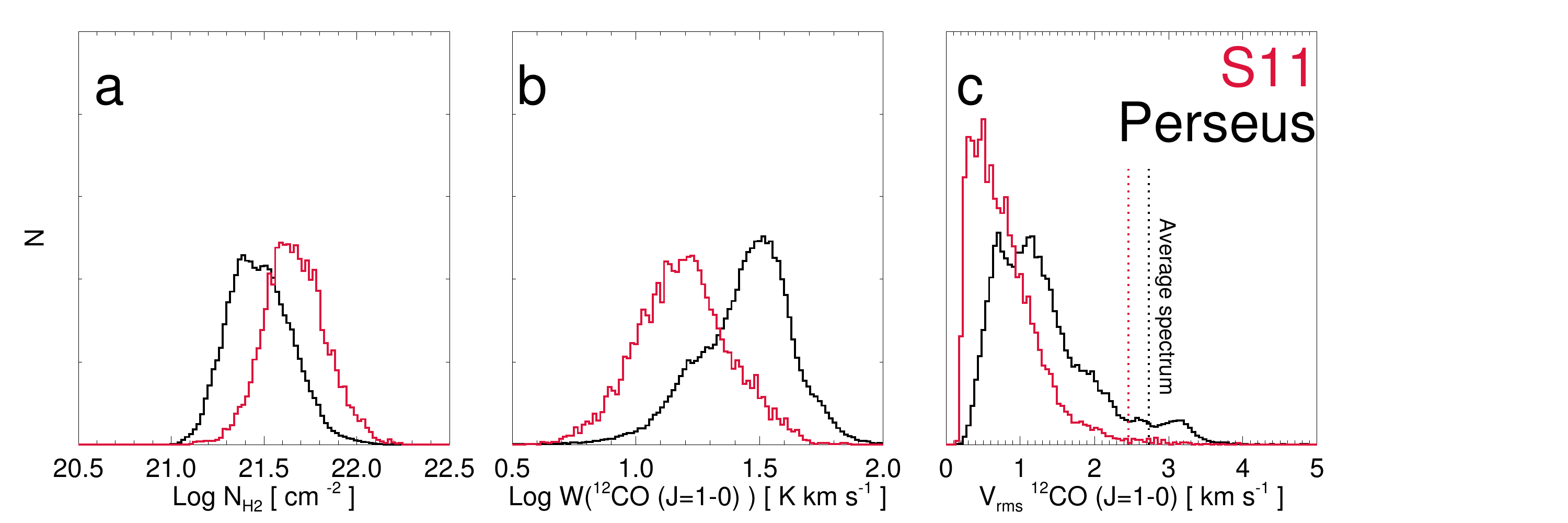}
\caption{Comparison of the line-of-sight distributions of column density (left), \coa\, integrated intensity, and \coa\, velocity dispersion for the S11 simulation (red) and Perseus (black). The dashed vertical lines show the velocity dispersion for the spatially-integrated cloud spectrum.}
\label{fig:diagnostic_rahul}
\end{figure*}

To make the differences between the simulations and Perseus more precise, Figure \ref{fig:diagnostic_rahul} shows three statistical comparisons between the S11 simulation and Perseus: the distribution of
column density, \coa~ integrated intensity, and line-of-sight velocity dispersion. The simulation has a higher average column density than Perseus and fainter CO lines. To rough approximation, integrated line intensity increases with gas density, temperature, and velocity dispersion. Since the S11 simulation is at a higher column density than Perseus (panel a), the stronger lines in Perseus are probably due to hotter gas, higher turbulence, and/or poor modeling of CO abundance.

The velocity dispersion of the spatially-averaged spectrum is shown as a vertical line in Figure \ref{fig:diagnostic_rahul}c. This number is larger than the typical line-of-sight velocity dispersion (the histograms in panel c), due to spatial velocity gradients across the region. In other words, the linewidth of the spatially-averaged spectrum is different from -- and larger than -- the mean of the line-of-sight linewidth distribution. While the spatially-averaged velocity dispersion in S11 is comparable to Perseus, the individual line-of-sight velocity dispersions in Perseus are skewed towards higher values. We
speculate that this is related to the characteristic depth of each line
of sight; lines of sight in S11 often intersect a single, $\sim$1pc-thick
 filament of material, which moves more coherently  -- i.e., with a
smaller velocity dispersion -- than the cloud as a whole. It may be that
typical lines of sight in Perseus pass through material spread out over
a longer column, and hence have larger dispersions. This may explain the higher integrated intensities in Perseus, since $W = \int T dv$

\begin{figure*}[htbp]
\centering
\includegraphics[width=7in]{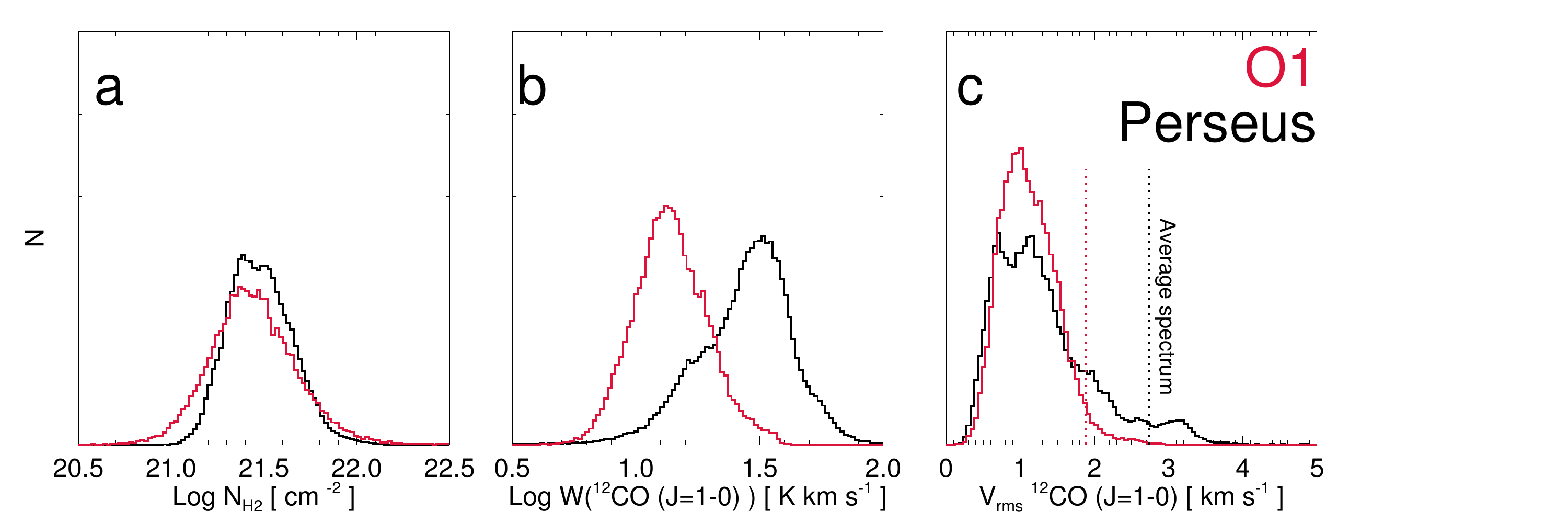}
\caption{Same as Figure 2, but with Simulation O1.}
\label{fig:diagnostic_stella}
\end{figure*}

The O1 simulation exhibits similar discrepancies (Figure \ref{fig:diagnostic_stella}). Its mean density and Mach number were
chosen to match Figures \ref{fig:diagnostic_stella}a and \ref{fig:diagnostic_stella}c. It better
reproduces the column density distribution in Perseus by construction, but like S11, the integrated emission is too faint. 
The mode of the line-of-sight velocity
dispersion matches Perseus moderately well (panel \ref{fig:diagnostic_stella}c), but Perseus has a longer
tail of high-velocity dispersion material. The velocity dispersion of the cloud-averaged spectrum (dashed lines in \ref{fig:diagnostic_stella}c) 
is 50\% larger in Perseus.

Turbulence in both simulations is arbitrarily driven, and this non-physical prescription does not reproduce statistics of the 1D line-of-sight velocity field in Perseus. One possibility for this discrepancy is that the simulations are not fully resolving the velocities on the smallest scales.  Resolution studies of grid based codes indicate that on small scales the amplitude of the velocity power spectrum is larger in simulations with higher resolution (e.g. \citealt{http://adsabs.harvard.edu/abs/2003ApJ...590..858V}).  For the intermediate resolutions considered here, therefore, the velocity dispersions on the smallest scales may be underestimated.  Additionally, some of the high-dispersion emission from Perseus coincides with the parsec-scale, stellar-driven shells studied in Arce et al. 2011 (in particular, Arce's CPS 4 and CPS 5). Neither simulation considers the effect of stellar feedback.  

Lee et al (2013) have also recently compared a spatially-truncated version of the S11 simulation to Perseus. They too note that the S11 simulation
is under-luminous. They report a more extreme under-luminosity in W(\coa) of 8x relative to Perseus, 
though their truncation acts to further diminish the line intensity. They posit that velocity crowding is responsible for this discrepancy --
the lower velocity dispersion in S11 makes superposition more likely, and optically thick $^{12}$CO features are more likely
to obscure each other. Indeed, we find that the peak of the W(\coc) distributions in both O1 and S11 better reproduce the values in Perseus (Appendix A).
The optical depth in W(\coc) is lower and should be less affected by velocity crowding. Nevertheless, Perseus still shows an excess of
large \coc\, intensities that neither simulation reproduces.

We hope that these discrepancies will motivate further efforts to
generate simulated clouds which better agree with the statistical
properties of Perseus. \textit{To that end, we discuss a suite of diagnostics in Appendix A}, 
to facilitate standardized comparisons in the future. For the purposes of this paper,
it is important to bear in mind that the ``observed'' properties in both
simulations are under-luminous and under-dispersed compared to
Perseus. 

\section{Results}
\label{sec:result}

\begin{figure}[htbp]
\centering
\includegraphics[width=3in, trim = 0 .53in 0 0, clip]{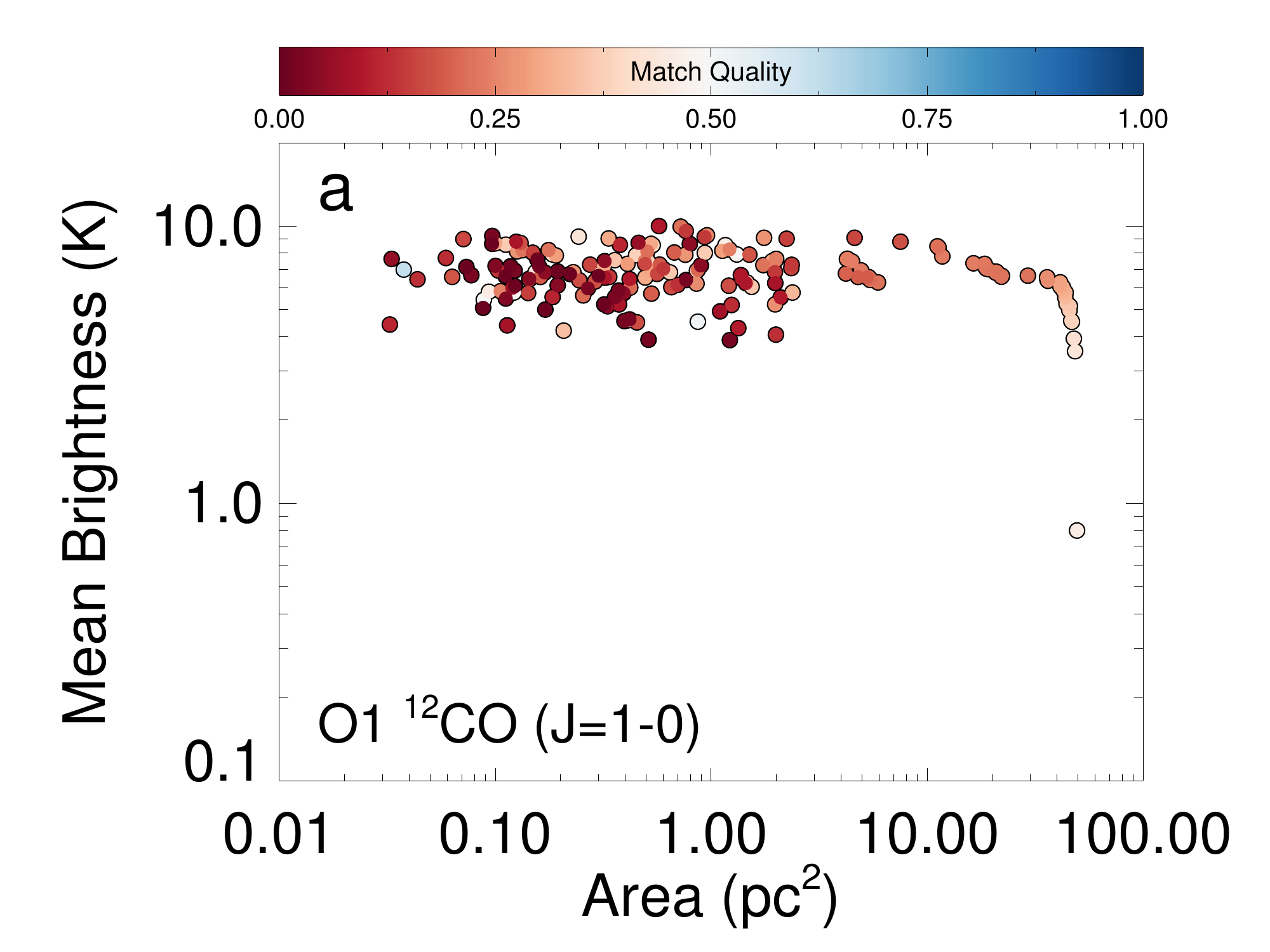}
\includegraphics[width=3in, trim = 0 .53in 0 .85in, clip]{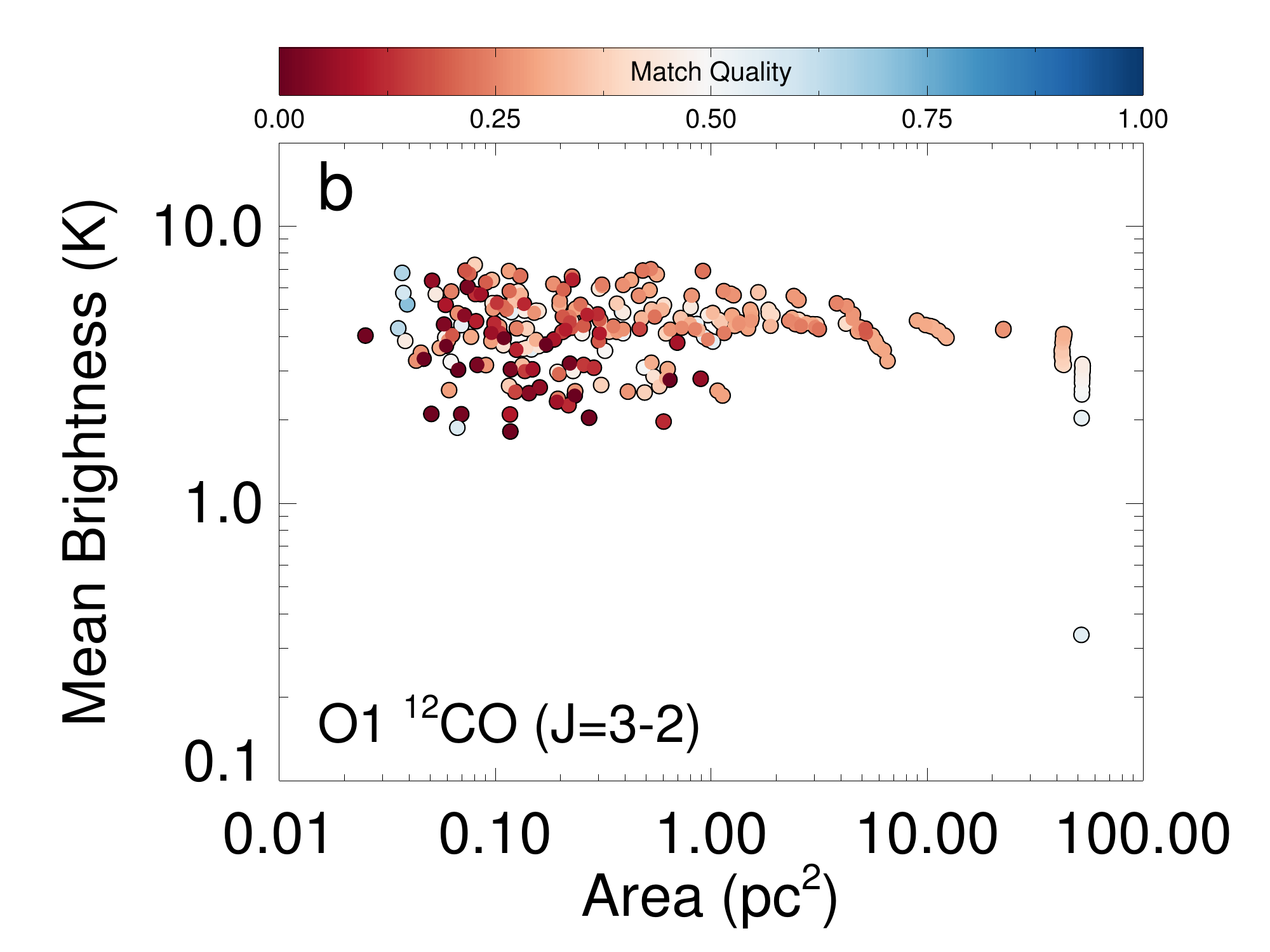}
\includegraphics[width=3in, trim = 0 0 0 .85in, clip]{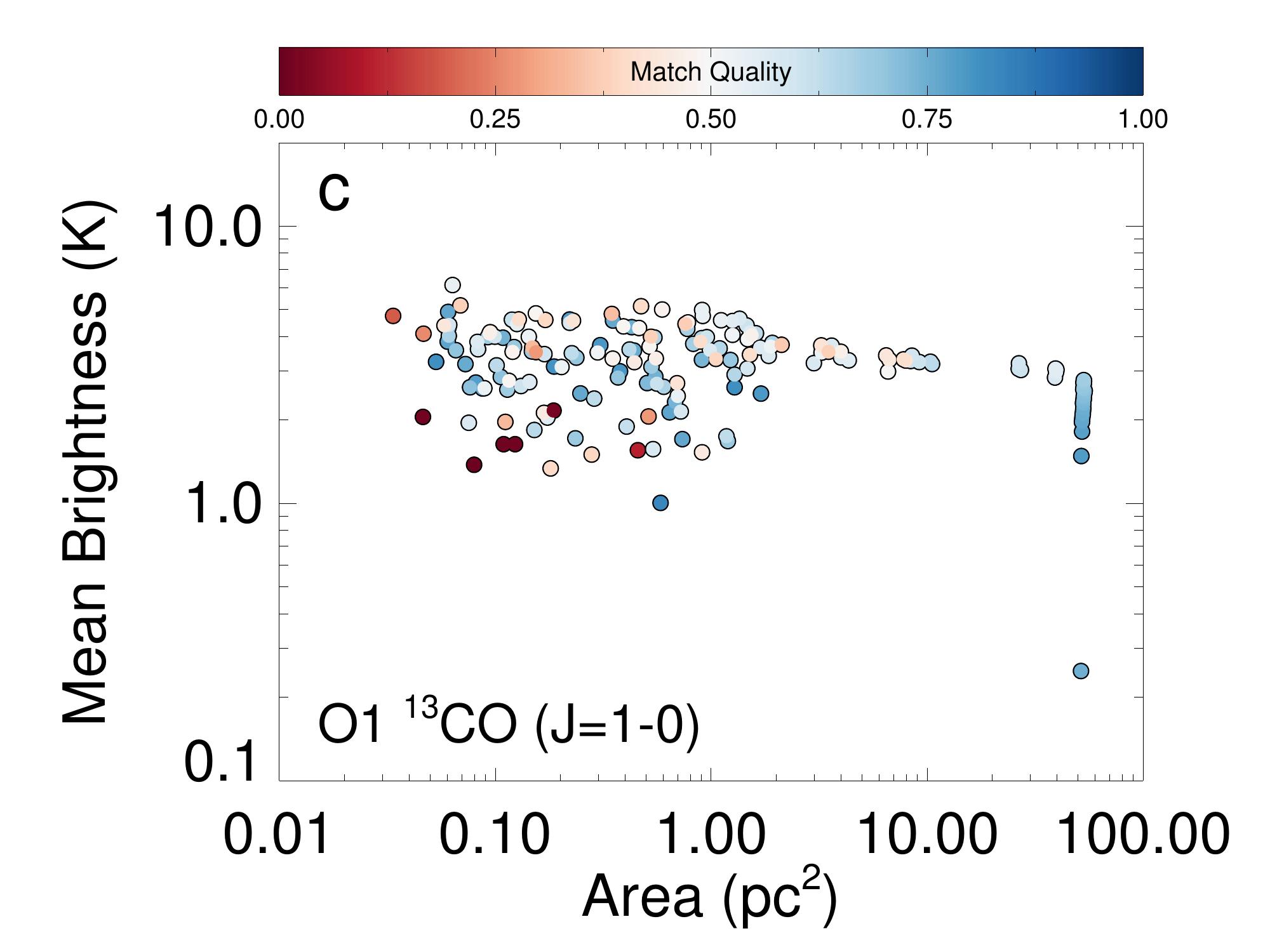}
\includegraphics[width=3in, trim = 0 0 0 .85in, clip]{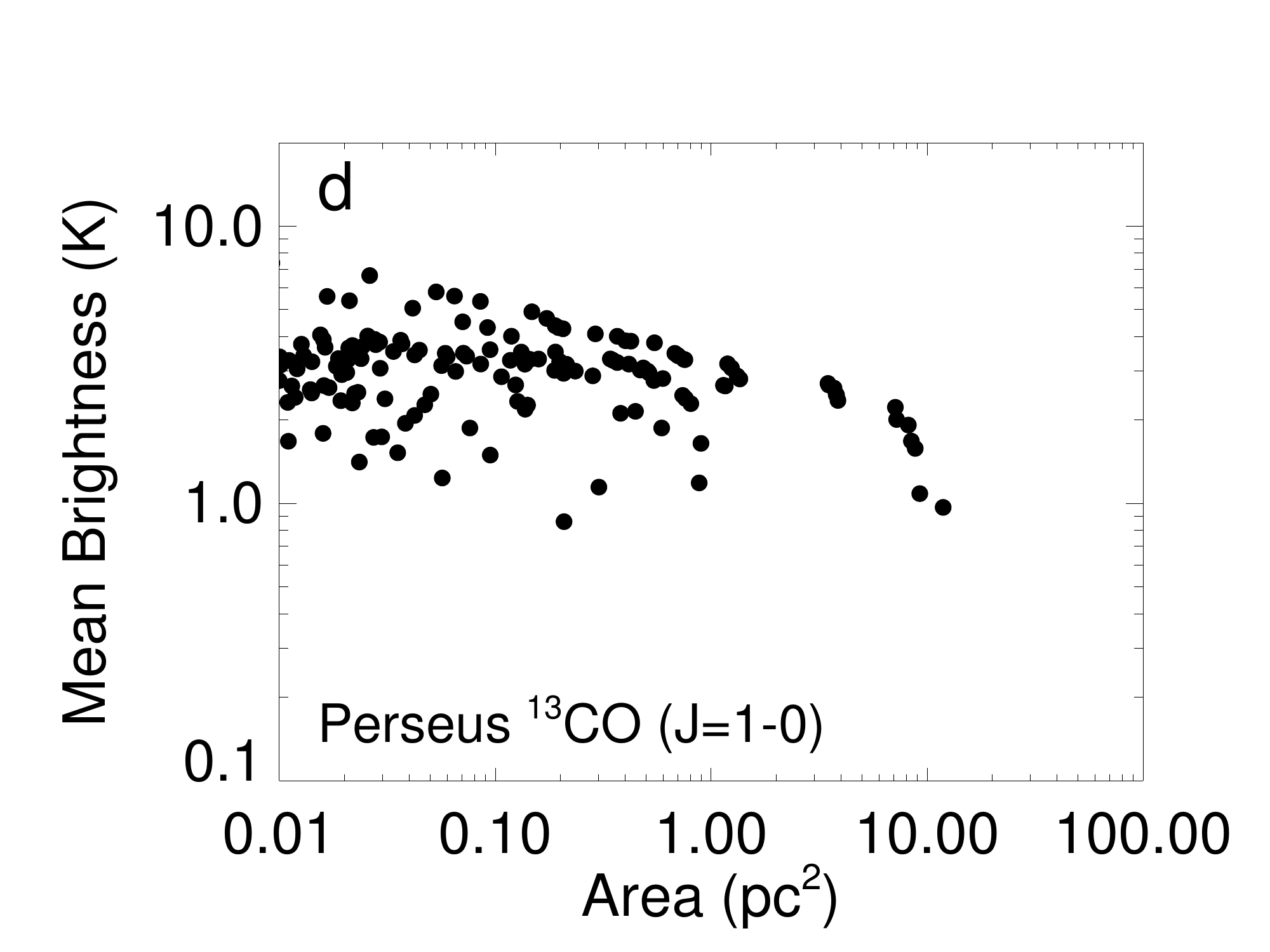}

\caption{Match similarity (color) as a function of structure size and mean intensity, for the three transitions in the O1 simulation.
Panel d shows the structure size and mean intensity for \coc\, structures in Perseus.}

\label{fig:sim_transition}
\end{figure}

\begin{figure*}[htbp]
\centering
\includegraphics[height=8in]{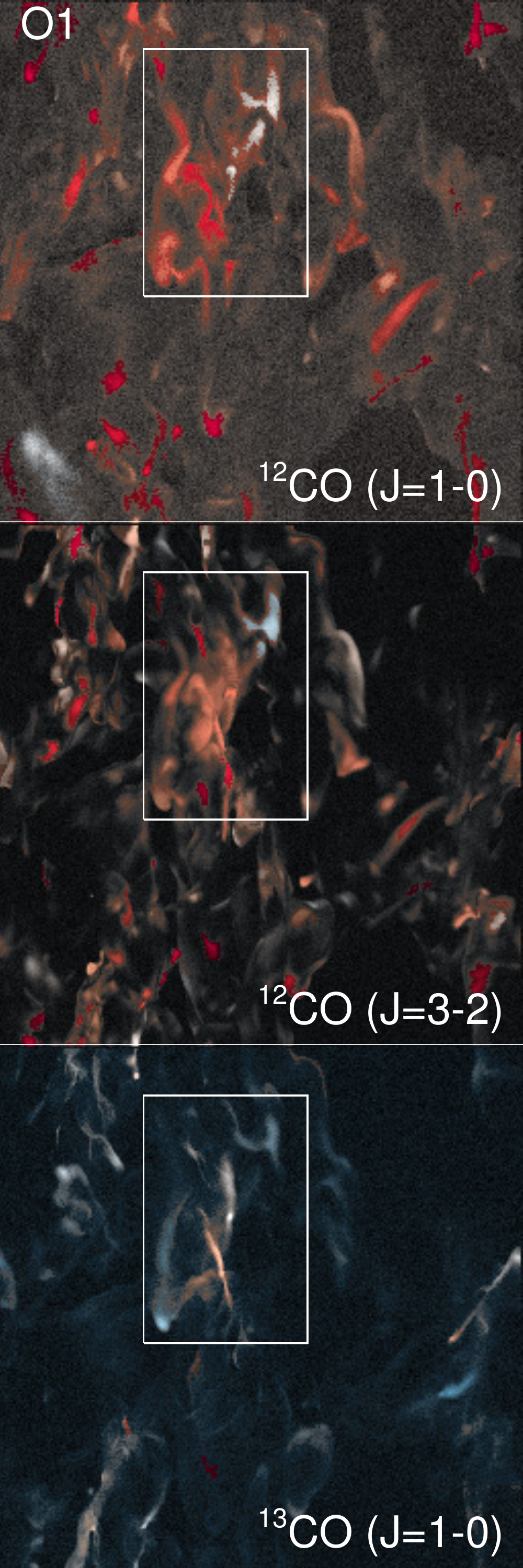}
\includegraphics[height=8in]{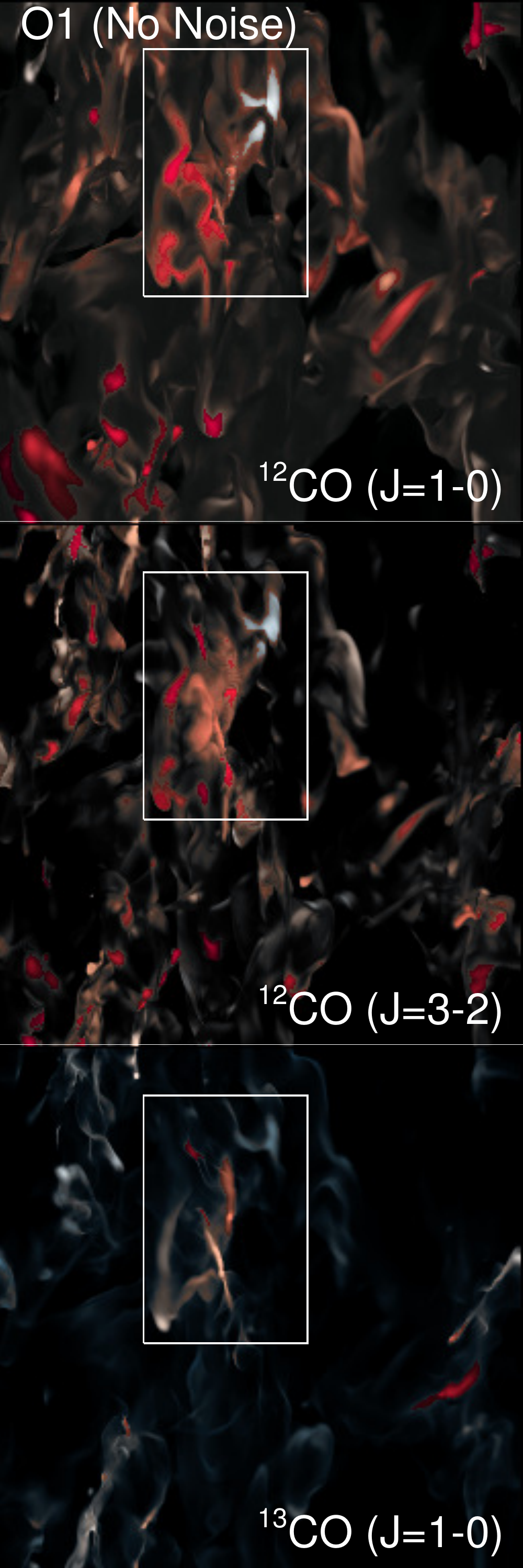}
\caption{Emission at a single velocity in each transition for O1, with (left) and without (right) noise, color-coded by match quality.}
\label{fig:im_stella}
\end{figure*}

For each simulation, we synthesize PPV observations of \coa, \cob, and \coc\, line emission.  Figure \ref{fig:sim_transition} shows
the match quality $q$ for features identified in each line observation of the O1 simulation. Each plot shows the $q$ (Equation \ref{eq:quality}) of a structure
(color) as a function of area and mean brightness. Several trends merit
discussion. Most obvious in Figure \ref{fig:sim_transition} is the variation amongst the three CO tracers. Structures identified in \coc\, are much
better representations of the underlying density field than \coa\, or \cob; \coc\, structures are dominated by structures with $q\geq 0.5$, which are rare in either of the more space-filling, optically thick transitions. We expect that in synthetic spectral-line maps of much higher-density-tracing species than CO, such as NH$_3$ or N$_2$H$^+$, overlap and superposition would be less of a problem. Similar trends in the dependence of $q$ on the CO transition can also be seen in Figures \ref{fig:im_stella} and \ref{fig:im_rahul}.

Second, there is a weaker trend between structure size and $q$ -- there is a left-to-right color gradient in Figure \ref{fig:sim_transition}a-c. Smaller structures tend to be more deeply embedded in cloud material, and are more susceptible to chance occlusion by or superposition with other structures. The large-scale features of the cloud, on the other hand, are less susceptible to superposition. Note that the simulations do not include other clouds along the line of sight; real clouds suffer confusion from other regions in the Galaxy. The dependence of $q$ on size is evident to various extents throughout Figures \ref{fig:sim_transition}-\ref{fig:virialvsvirial_nochem}.

Finally and most subtly, low quality points tend to cluster towards smaller brightnesses at a given scale -- the lowest-quality points in Figure \ref{fig:sim_transition}b-c are skewed towards the lower envelope of points. As discussed above, low-quality structures
correspond to superposition artifacts, or pseudo- (i.e.~non-density)
structures created by radiative transfer effects or spatial variation in
the $v_z$ field. Because of this, the size and intensity of artifacts
depend on how `organized' these processes are; smaller and fainter artifacts are more probable since, in the case of superposition,
they require only a partial overlap of two real structures or, in the
case of velocity-induced structures, only a small scale
organization in $v_z$.

Figure \ref{fig:im_stella} shows a single PP slice in the O1
simulation. Again, the color scale gives the match quality for each
structure\footnote{Due to the hierarchical nature of dendrogram decomposition, a single location is generally associated with multiple nested structures. In these images, each pixel is colored according to the smallest and densest feature to which it belongs. Essentially, these figures show the ``worst-case'' scenario view, since larger structures typically offer better matches.}. The interior box draws attention to one particularly crowded region. The \coa\, simulation
is most affected by confusion in this region, and has a lower average match quality.  Line saturation tends to broaden
features in \coa such that the morphology in that transition
is less representative of the true density field which, as the \coc\, transition suggests, is more compact.

\subsection{O1 vs S11}
\begin{figure}[htbp]
\includegraphics[width=3in]{stella_dss_co10}
\includegraphics[width=3in]{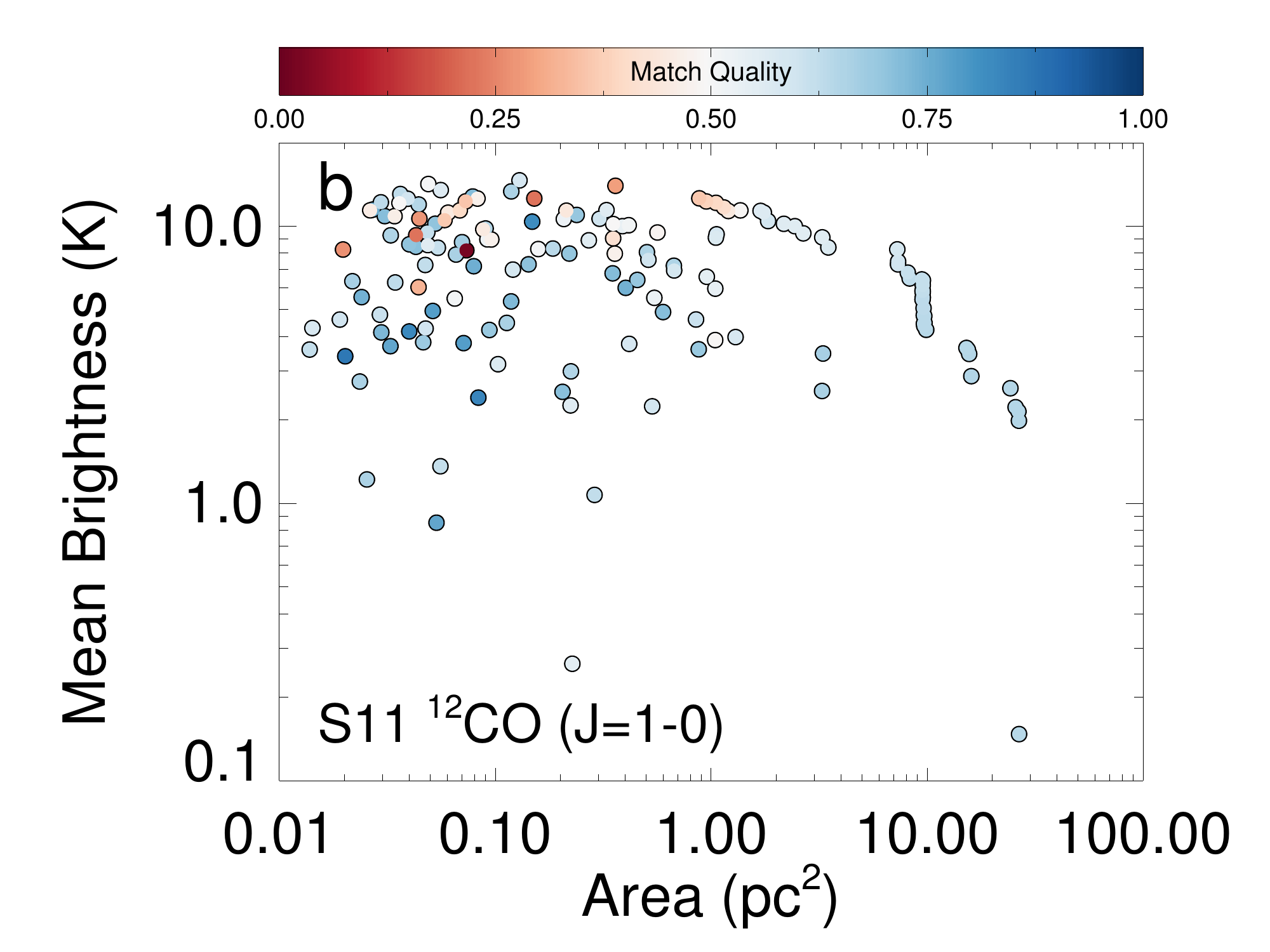}
\caption{The same as Figure \ref{fig:sim_transition}, but comparing the \coa\, transition in the S11 and O1 simulations}
\label{fig:sim_simulation}
\end{figure}

\begin{figure}[htbp]
\centering
\includegraphics[height=8in]{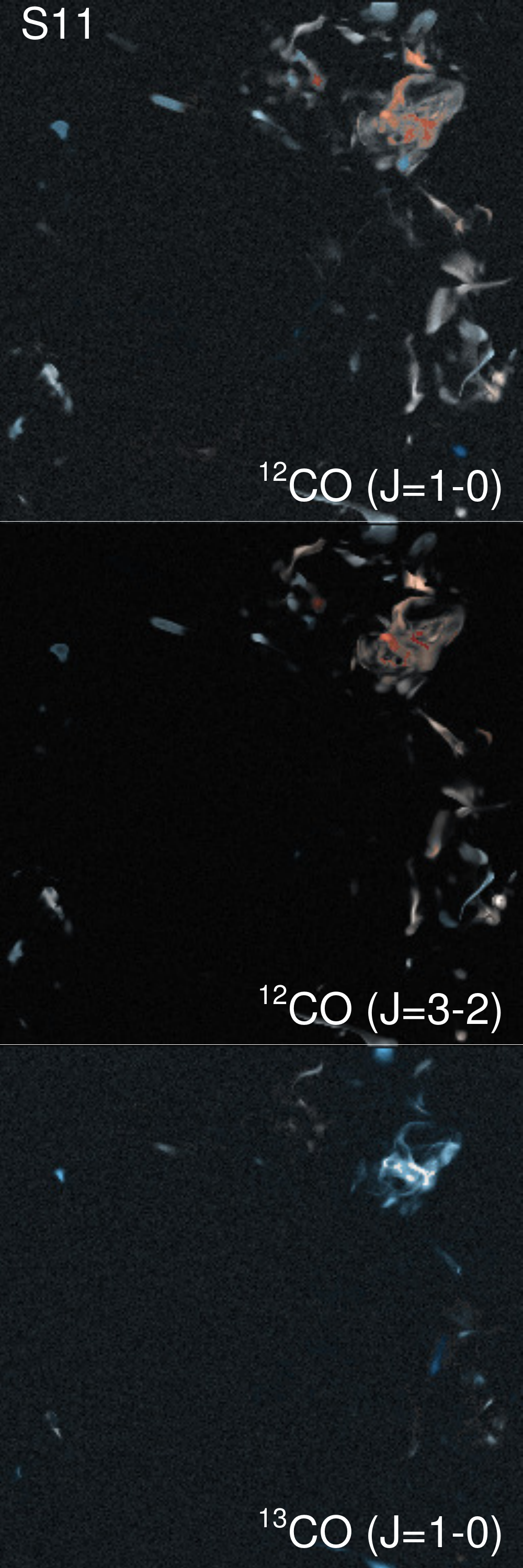}
\caption{The same as Figure \ref{fig:im_stella}, but for the S11 simulation}
\label{fig:im_rahul}
\end{figure}

As a first comparison between the properties of the O1 and S11 simulations, Figure \ref{fig:sim_simulation} compares the \coa\, transitions for each simulation, and Figure \ref{fig:im_rahul} shows a color-coded PP slice of the S11 simulation. The O1 and S11 simulations show marked differences. The S11 simulation has overall better match quality. The S11 simulation also has a
higher dynamic range of structure brightnesses; this is probably due to
the fact that that simulation explicitly treated gas heating and
CO dissociation, whereas the O1 simulation is isothermal and assumes a constant CO abundance. Heating and dissociation
give the S11 simulation more freedom to affect line intensity (by raising the excitation temperature as well as
the abundance of emitting material). CO is dissociated in low column density regions of S11, and the simulation has large regions devoid of emission.

\subsection{Effect of Noise}
Noise has been added to each synthetic observation, to match the noise levels in present-day cloud
observations in these transitions (Section \ref{sec:sim}). This raises the following question: to what extent does noise make it more difficult to extract cloud features, and hence lower the match quality? To address this, we also show in Figure \ref{fig:im_stella} the
same PP slices without noise. Note that the presence or absence of noise
has little bearing on the match quality of most structures.  Instead, the lower match quality in the \coa\, transition seems dominated by
the high filling factor and opacity of emission, which crowds features in PPV and leads to superposition.

\subsection{Disentangling Projection and Radiative Transfer Effects}
The ability to recover structures varies with tracer; Figure \ref{fig:sim_transition} shows that structures in \coa\, 
are most affected by projection problems, followed by \cob\, and \coc. The latter two transitions trace higher densities
and are optically thinner than \coa. It is unclear from Figure \ref{fig:sim_transition} alone whether the poor match quality in \coa\, is the result of the higher filling factor
or higher opacity in that line.

We can partially decouple the effects of filling factor and opacity by disabling absorption in the radiative transfer. This doesn't prevent crowding or superposition in PPV, but it does prevent structures from blocking radiation. If opacity is the primary problem
with \coa\, observations, then disabling absorption should increase the overall match quality.

To test the effects of filling factor alone, we perform a modified radiative transfer calculation on the O1 simulation, where opacity is disabled. The full
equation of radiative transfer is given by

\begin{equation}
I_\nu = \int e^{-\tau(z)} B_\nu\left(T_{\rm ex}\left(z\right)\right) \alpha(z)\, \rm dz
\label{eq:rt_full}
\end{equation}
Where $\tau(z)$ is the optical depth to depth $z$, $B_\nu$ is the Planck function, $T_{\rm ex}(z)$ is the excitation
temperature of the gas, and $\alpha$ the absorption coefficient.  $T_{ex}$ and $\alpha$ are functions of the gas density at each energy level. 

For our modified radiative transfer calculation, we run RADMC-3D as normal to compute the level populations and hence $T_{\rm ex}$ and $\alpha$
throughout the simulation volume. However, we then integrate a modified equation of radiative transfer with no absorption:

\begin{equation}
\tilde{I}_\nu = \int B_\nu\left(T_{\rm ex}\left(z\right)\right) \alpha(z)\, \rm dz
\end{equation}

\begin{figure}[htbp]
\centering
\includegraphics[width=3in]{stella_dss_co10}
\includegraphics[width=3in]{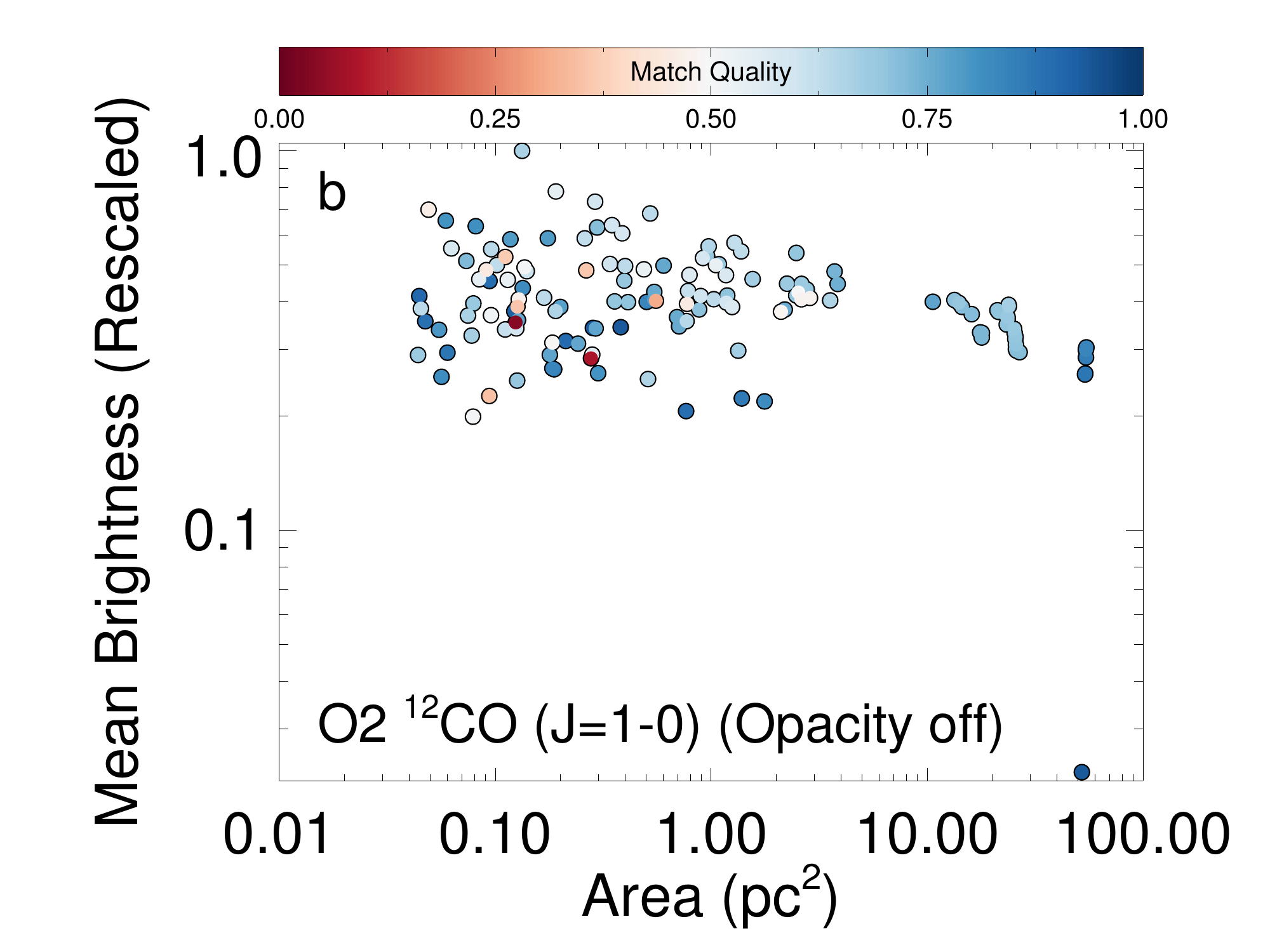}
\caption{The same as Figure \ref{fig:sim_transition}, but for the O2 simulation where opacity 
was disabled during radiative transfer.}
\label{fig:sim_notau}
\end{figure}

This produces a modified version of the O1 synthetic observations, which we label as O2. The only difference between O1 and O2 is that O2 includes no absorption. Figure \ref{fig:sim_notau} compares
the match quality for the \coa\, transition in simulations O1 and O2. The O2 structures are markedly higher quality. Figure \ref{fig:sim_notau} indicts the $e^{-\tau(z)}$ term in Equation \ref{eq:rt_full} as a primary reason for low match qualities in the \coa\, transition. 

Note that this experiment does not fully disable the effects of opacity. In addition to the $e^{-\tau(z)}$ term, opacity acts to increase the excitation temperature of the gas by absorbing radiation emitted by other parts of the cloud. Disabling this absorption could lead to de-excitation of parts of the cloud, and this spatially-varying excitation would partially decouple the intensity field from the density field and decrease the filling factor of emission.

\subsection{Effect of Pruning}
\label{sec:pruning_test}

\begin{figure}[htbp]
\centering
\includegraphics[width=3.4in]{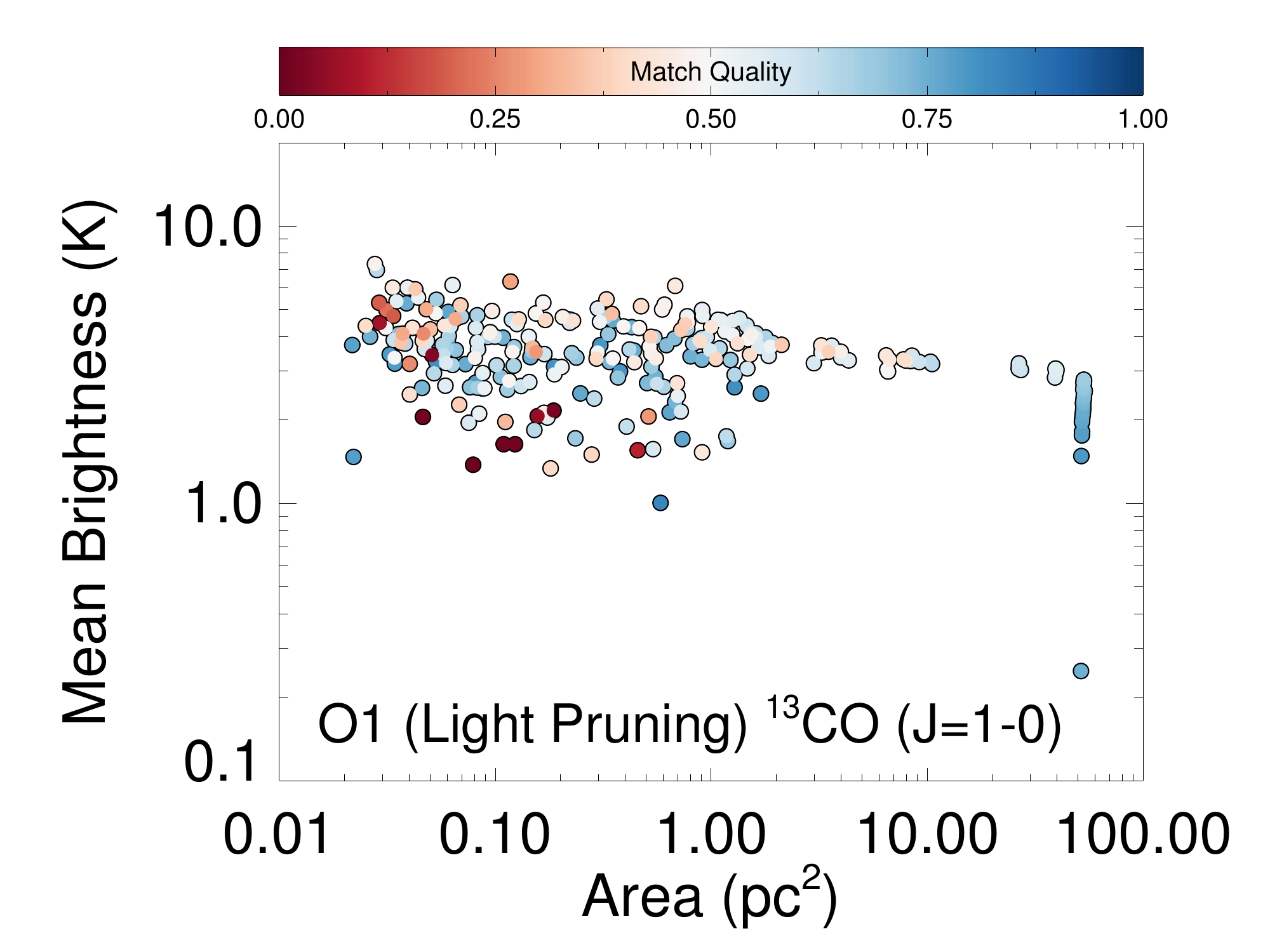}
\caption{The same as Figure \ref{fig:sim_transition}, but for the O1 simulation with less pruning.}
\label{fig:sim_prune}
\end{figure}

In Section \ref{sec:structure_identification}, we described our pruning strategy -- namely, we require that each structure contains a voxel 7$\sigma$ above the ambient intensity and contains at least 800 voxels altogether. Figure \ref{fig:sim_prune} shows a less aggressive pruning strategy, where we relax the $N>800$ voxel criterion to $N>400$. There are more structures in this dendrogram (283 structures in the \coc\, transition, compared to 191 in the original pruning). Compared to Figure \ref{fig:sim_transition}c, these additional points are concentrated at Areas $< 0.5$ pc$^2$. We reiterate that the points in Figure \ref{fig:sim_prune} are a superset of Figure \ref{fig:sim_transition}c, and the extra points are substructures nested inside the structures
from the original pruning.

 On average, the new structures have modestly lower match qualities: 38\% of the new structures have $q<0.5$, compared to 27\% of structures in the original pruning. However, we interpret Figure \ref{fig:sim_prune} as evidence that the reality or quality of dendrogram-identified structures is fairly insensitive to the details of pruning, provided that statistically-insignificant noise-spikes are not identified as structures.

\subsection{Impact on Scaling Relations}

What impact does confusion caused by projection and radiative transfer
have on subsequent analyses? This is a problem-dependent question,
and we focus here on the fairly common virial analysis. The virial
parameter, often defined as $\alpha= 5 \sigma_v^2 R / G M $, gives
the approximate ratio of kinetic to gravitational potential energy
\citep{http://adsabs.harvard.edu/abs/1992ApJ...399..551M}. The value
$\alpha \sim 2 $ denotes the approximate equipartition between these two
energy terms and is often used to assess the boundedness of a given
structure. However, the true virial state of an object is affected by 
several additional unobservable terms (for example, surface terms and magnetic energy;
\citealt{http://adsabs.harvard.edu/abs/2006MNRAS.372..443B, http://adsabs.harvard.edu/abs/2007ApJ...661..262D, http://adsabs.harvard.edu/abs/1992ApJ...395..140B}), and we
emphasize that the $\alpha < 2$ threshold is a crude proxy for
boundedness. Furthermore, the virial analysis implicitly assumes that structures are roughly
spherically-symmetric, which does not well-describe the larger features in a dendrogram of a filamentary cloud.

\begin{figure*}[htbp]
\includegraphics[width=6in]{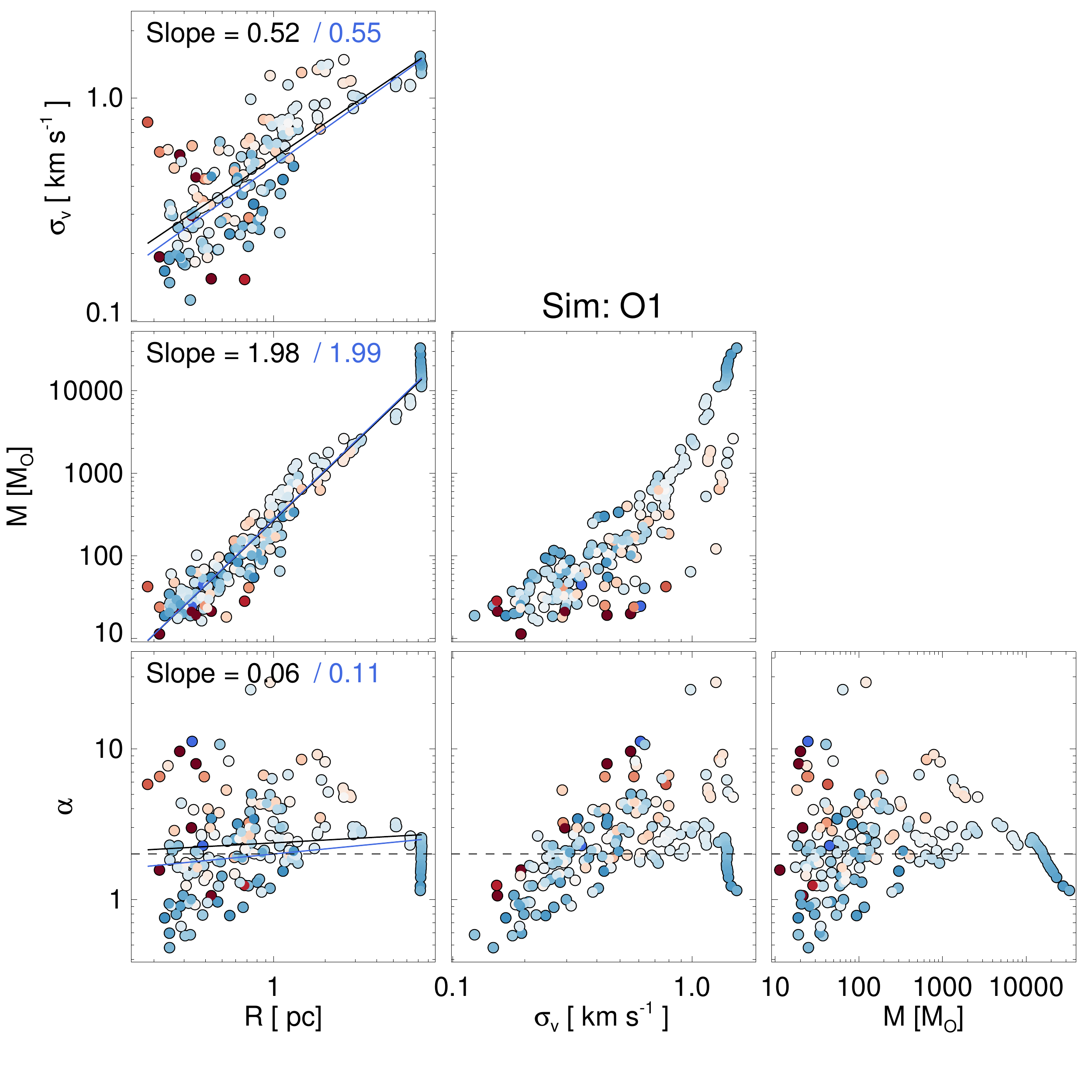}
\caption{Scatter matrix of mass, size, velocity dispersion, and virial parameter for the \coc\, transition of simulation O1. 
Points are color-coded by match quality. The black line is the linear fit to all points, while the blue line is for $q > 0.5$.}
\label{fig:scatter_matrix}
\end{figure*}

Figure \ref{fig:scatter_matrix} shows, for the \coc\, transition of simulation O1, the relationship between size, velocity dispersion, mass, and
virial parameter for each structure in the dendrogram. Each point is color-coded by match quality as before.  Appendix B describes how each quantity is
measured. 

We also show a simple power law fit to the size-linewidth, mass-size, and virial-size relationship, which are frequently measured in molecular cloud studies. The black line shows the fit to all points, while the blue line shows structures with $q > 0.5$. The slopes of these lines essentially reproduce Larson's classical relationships ($M \sim R^2$, $V_{\rm rms}\sim R^{0.5}$, $\alpha \sim R^0$; \citealt{http://adsabs.harvard.edu/abs/1981MNRAS.194..809L}). Ignoring the low-quality match structures affects the slope by $\lesssim 0.05$.

Assessing the uncertainty in these scaling relationships is subtle. A naive least-squares error analysis suggests a small uncertainty for the scaling exponents ($\sim .025$). However, these data points correspond to nested structures and are not independent of each other. Consequently, the least-squares error estimate is overly optimistic. We have experimented with different sensible strategies for pruning the dendrogram, as well as different definitions for the size of an irregular structure (see Appendix B). Varying these options can change the slope of the scaling relationships by $\sim \pm 0.2$ (see, for example, Table \ref{tab:scale_scatter}), and we feel this is a more appropriate estimate for how precisely the scaling relationships are constrained.

\begin{deluxetable}{lr}
\tablecolumns{2}
\tablewidth{0in}
\tabletypesize{\scriptsize}
\tablecaption{Sensitivity of mass-size relationship to size definition}
\tablehead{\colhead{Size definition} & \colhead{Scaling coefficient $M \sim r^{a}$}}
\startdata
Second moment\tablenotemark{a,b} & 0.48 \\ 
$r = V^{1/3}$ & 0.34 \\
$r = A^{1/2}$ \tablenotemark{c} & 0.43 \\
\enddata
\tablenotetext{a}{Value obtained via least squares fit to the O1 data}
\tablenotetext{b}{The size definition used throughout this paper and described in Appendix B.}
\tablenotetext{c}{$A$ is the area of the structure projected onto the sky.}
\label{tab:scale_scatter}
\end{deluxetable}

For the O1 simulation, then, filtering based on $q$ does not significantly affect the scaling relationships one obtains from PPV data. 

\subsection{Parameters Compared as Measured in PPP and PPV}

PPV-derived properties are most often used as approximations for properties of the (partially un-measurable) 6-dimensional spatial-kinematic state of the cloud. How accurate are these approximations?

\begin{figure*}[htbp]
\includegraphics[width=3.4in]{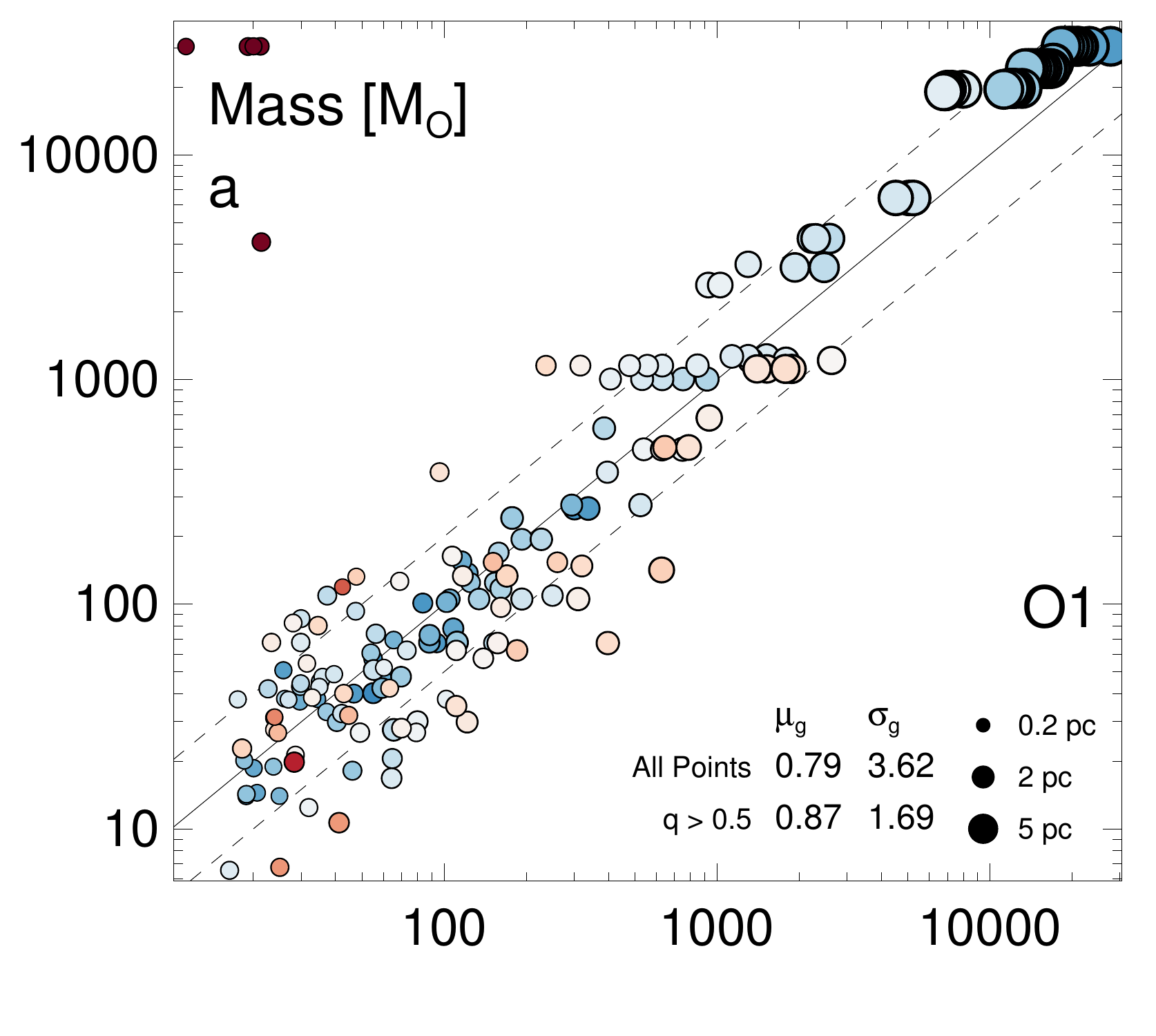}
\hfill
\includegraphics[width=3.4in]{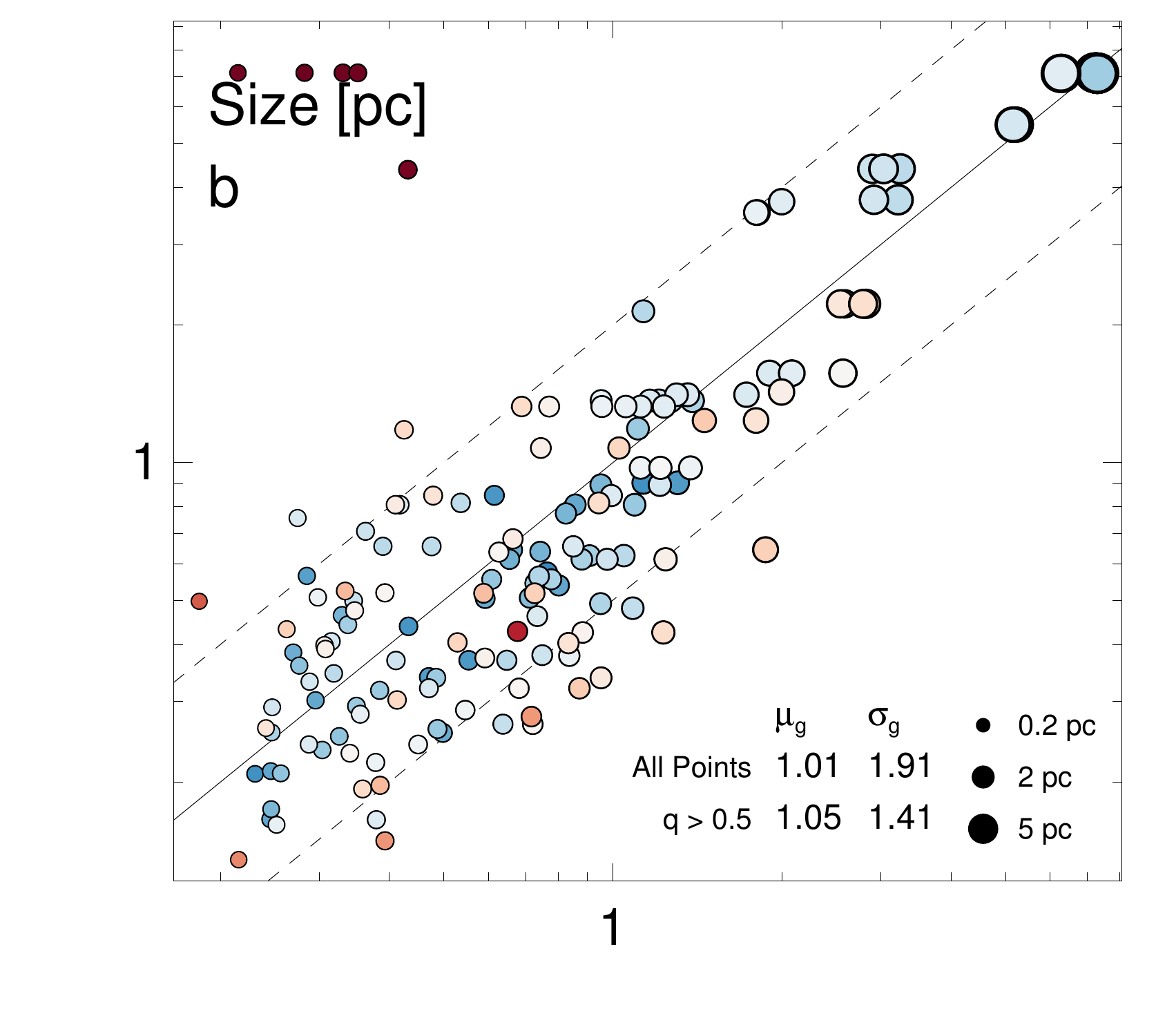}
\includegraphics[width=3.4in]{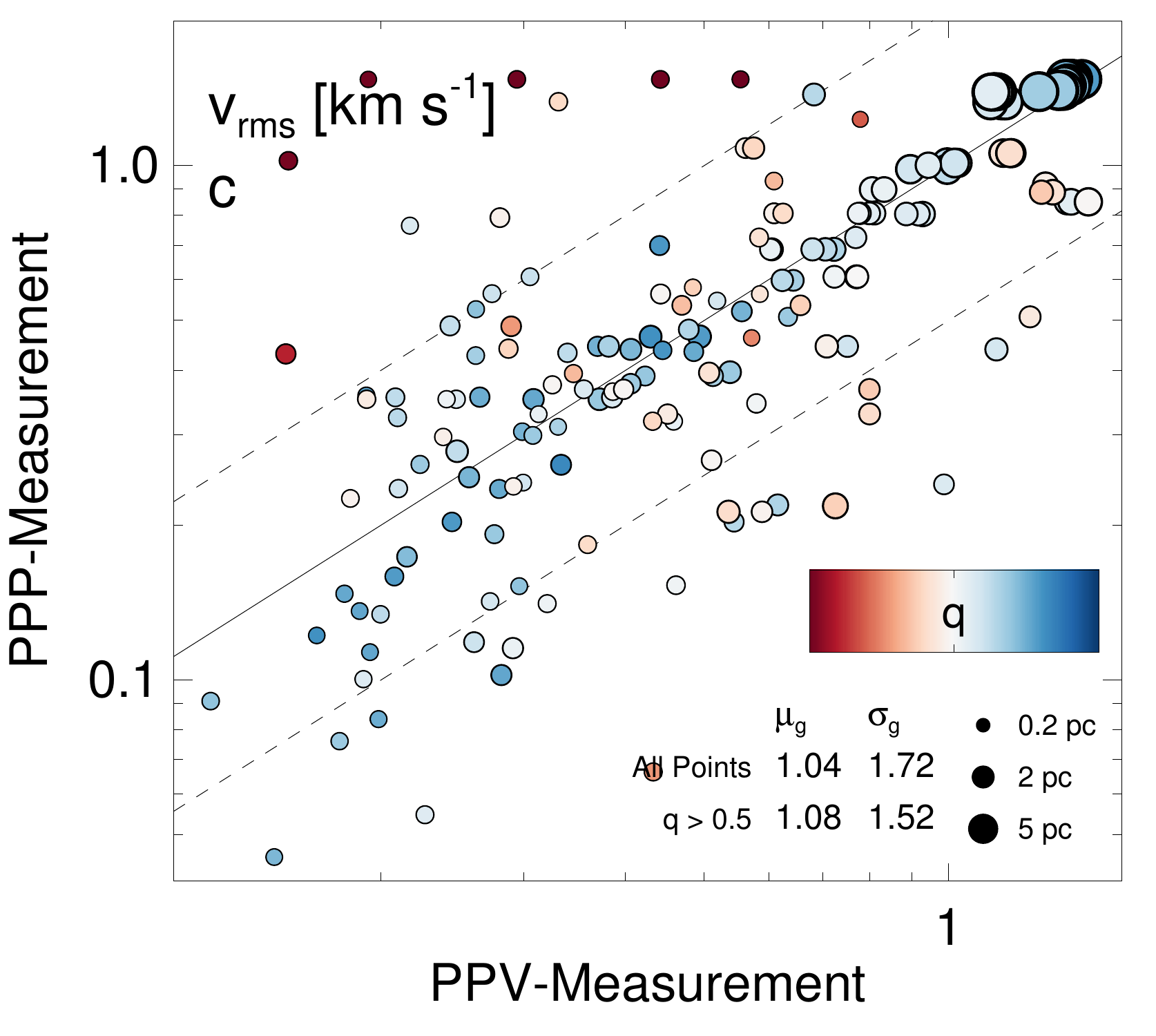}
\hfill
\includegraphics[width=3.4in]{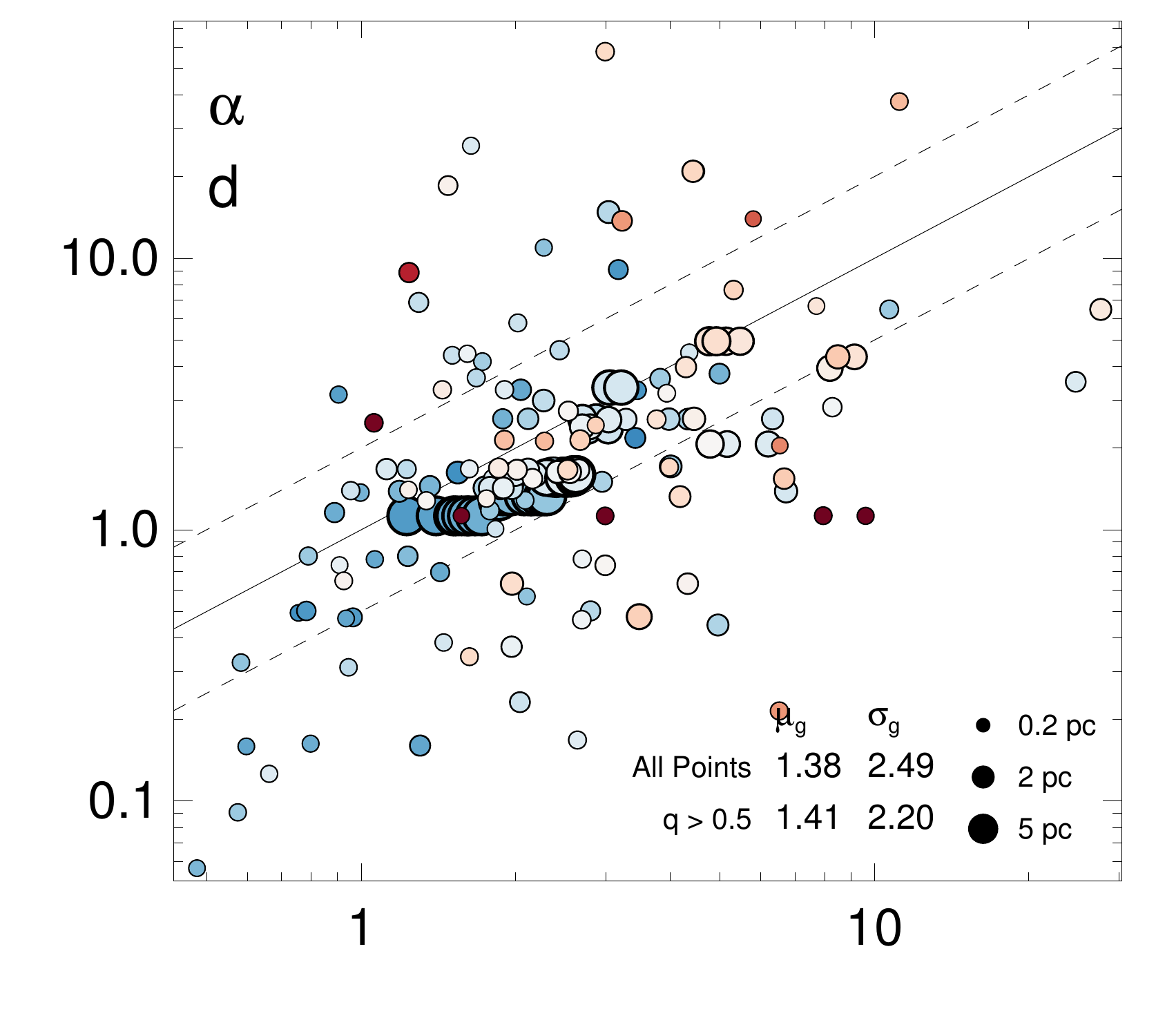}
\caption{A comparison between mass, size, linewidth, and virial parameter measurements (panels a, b, c, d) derived from PPV and PPP data, for the \coc\, transition of O1. Points are colored by match quality, and scaled by structure size. The dashed lines are a factor of two above/below the solid 1:1 line.}
\label{fig:virialvsvirial}
\end{figure*}

Figure \ref{fig:virialvsvirial} compares, for the \coc\, transition of O1, quantities measured in PPV with the equivalent measurement of each PPV structure's nearest PPP match (measured in PPP). The point sizes indicate structure size, and the dashed lines are a factor of 
2 above and below the 1:1 line. In the lower right corner of each panel, we also report a few summary statistics: the geometric mean of the ratio of PPV/PPP measurements and the geometric standard deviation of this ratio. The geometric mean is defined as $\mu_g = \left(\prod{x_i}\right)^{\frac1{N}}$ and the geometric standard deviation as $\sigma_g = \exp{\left(\sqrt{\frac1{N} \sum{\ln(x_i / \mu_g)}}\right)}$. The base-10 log of $\sigma_g$ is the scatter about the ratio $\mu_g$ in dex. These numbers measure the fractional bias from and scatter about the 1:1 line, respectively. We report the geometric mean and standard deviation for all points, as well as the subset of points with $q > 0.5$. In this analysis, we exclude structures which touch the edge of the cube, since their full extent is not measured.

There are several features of Figure \ref{fig:virialvsvirial} worth commenting on. First, the strongest outliers are the reddest, lowest-quality PPV structures, shown as the red points in the upper-left corner of panels a-c. These structures have no correspondence to any PPP structure. Because these artificial features cannot be sensibly matched to anything in the PPP cube, they are arbitrarily matched to the largest density structures and occupy the upper left corners of panel a-c. 

Second, structures with $q>0.5$ are dispersed about the 1:1 line by $\sigma_g=1.4-1.7$ ($\sim 0.1-0.2$ dex) in panels a-c. The act of filtering on $q$ reduces scatter by 15-30\% for mass and velocity. The reduction is larger for masses in panel a, but the dispersion in these points is dominated by the handful of outliers in the upper left corner.

Finally, the virial parameter plot (panel d) shows higher scatter for $q > 0.5$ structures  -- 0.34 dex, or a  factor of 2.2 -- than do the mass, size, or linewidth plots. The individual errors in the mass, size, and velocity dispersion measurements compound when measuring the virial parameter, and this property is the least-faithfully recovered. 

To summarize Figure \ref{fig:virialvsvirial}, radiative transfer and projection effects produce a factor of $1.4-2$ uncertainty on kinematic properties derived from the O1 simulation using \coc\, emission.

\subsection{Effect of Gravity}

\begin{figure*}[htbp]
\includegraphics[width=3.4in]{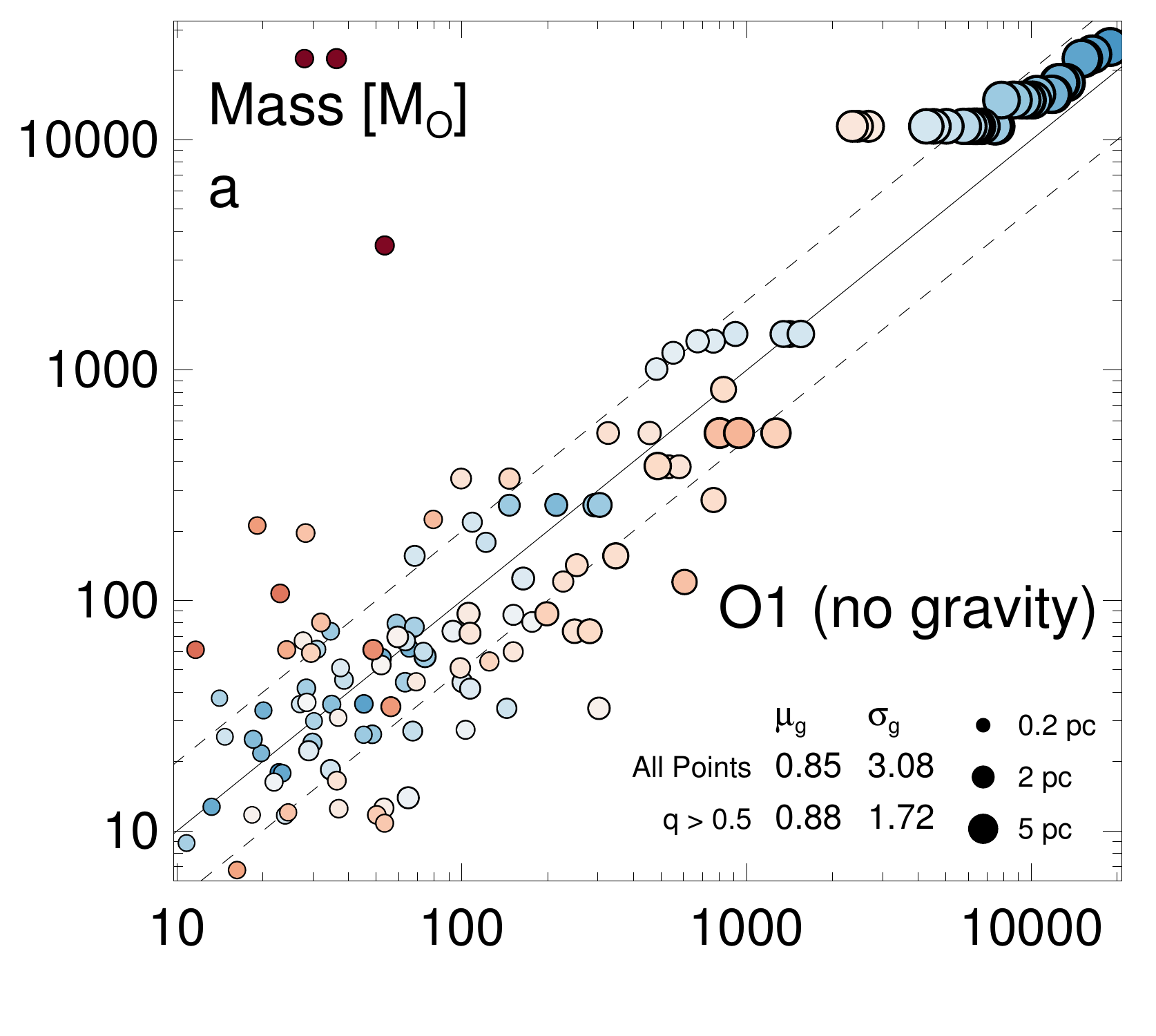}
\hfill
\includegraphics[width=3.4in]{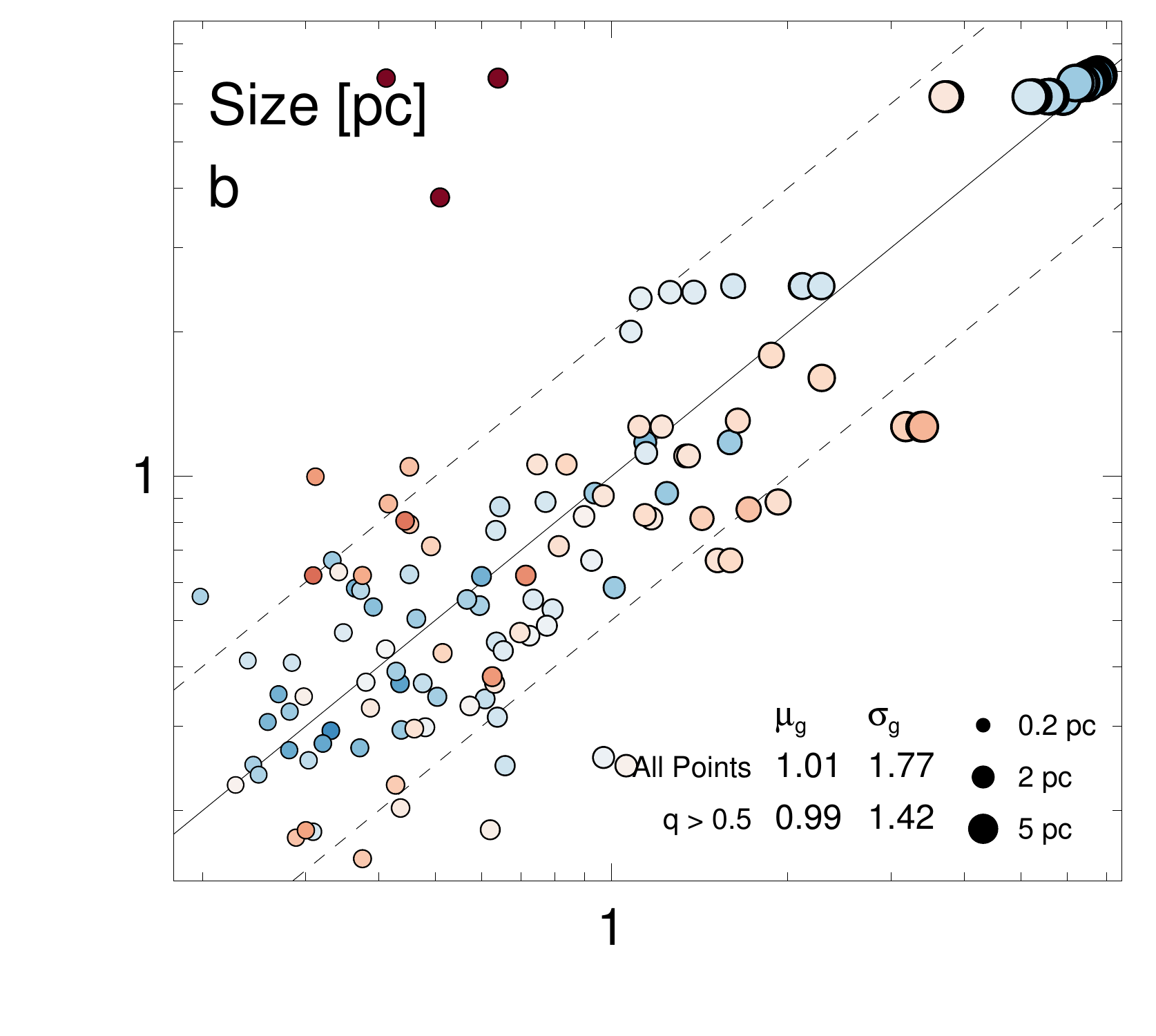}
\includegraphics[width=3.4in]{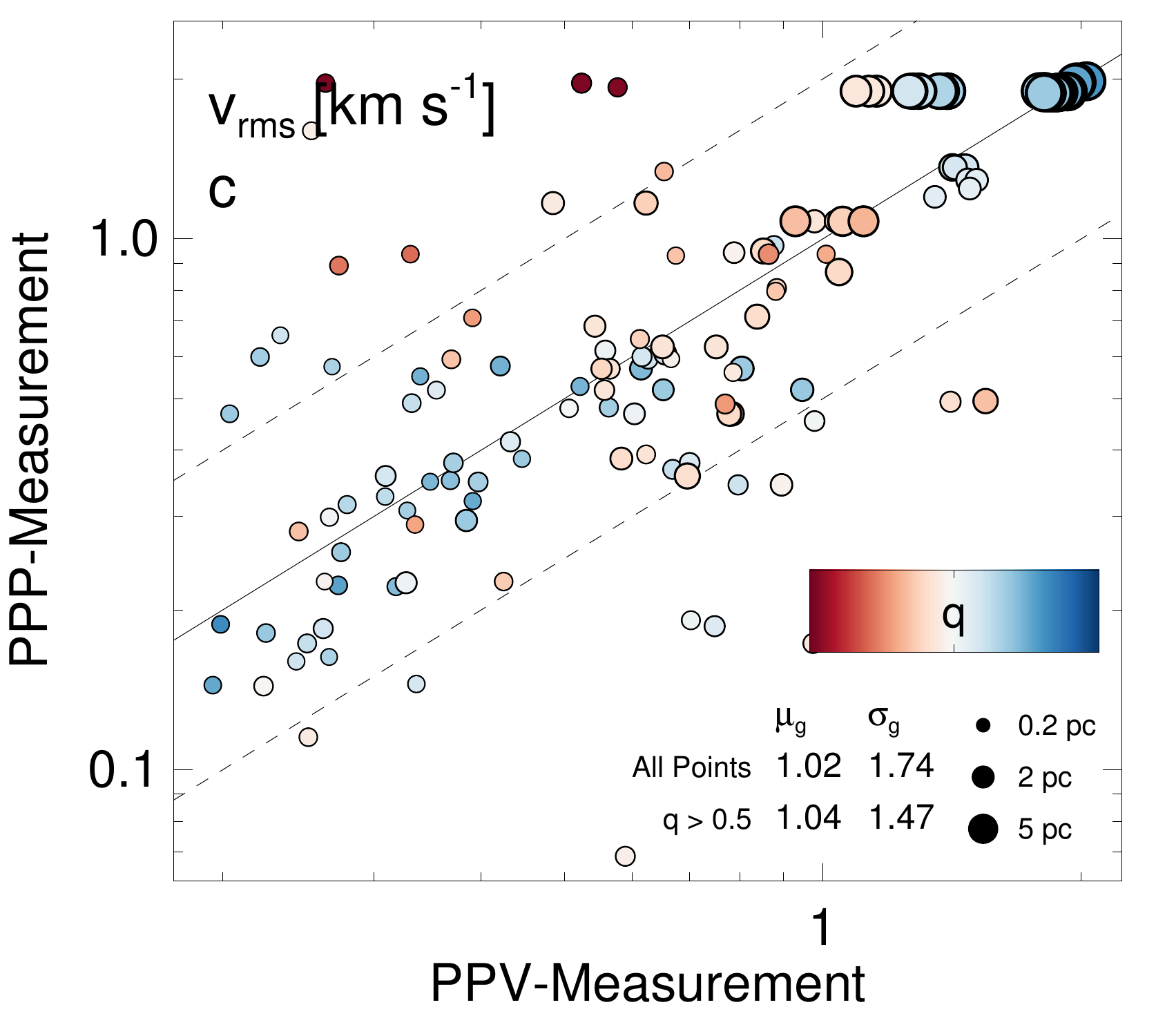}
\hfill
\includegraphics[width=3.4in]{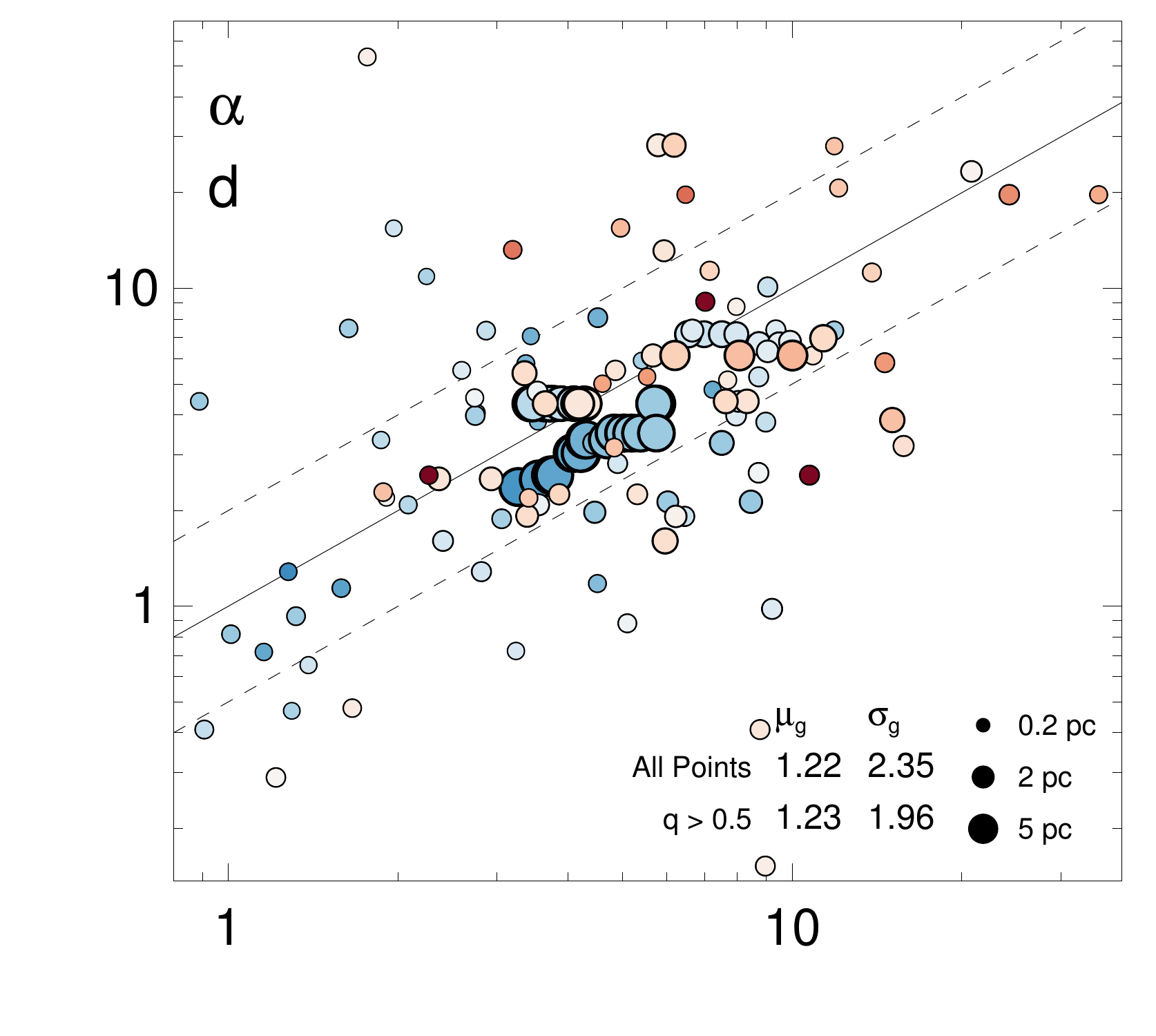}
\caption{The same as Figure \ref{fig:virialvsvirial}, but for the O1 simulation with no gravity.}
\label{fig:virialvsvirial_nog}
\end{figure*}

\begin{figure*}[htbp]
\includegraphics[width=6in]{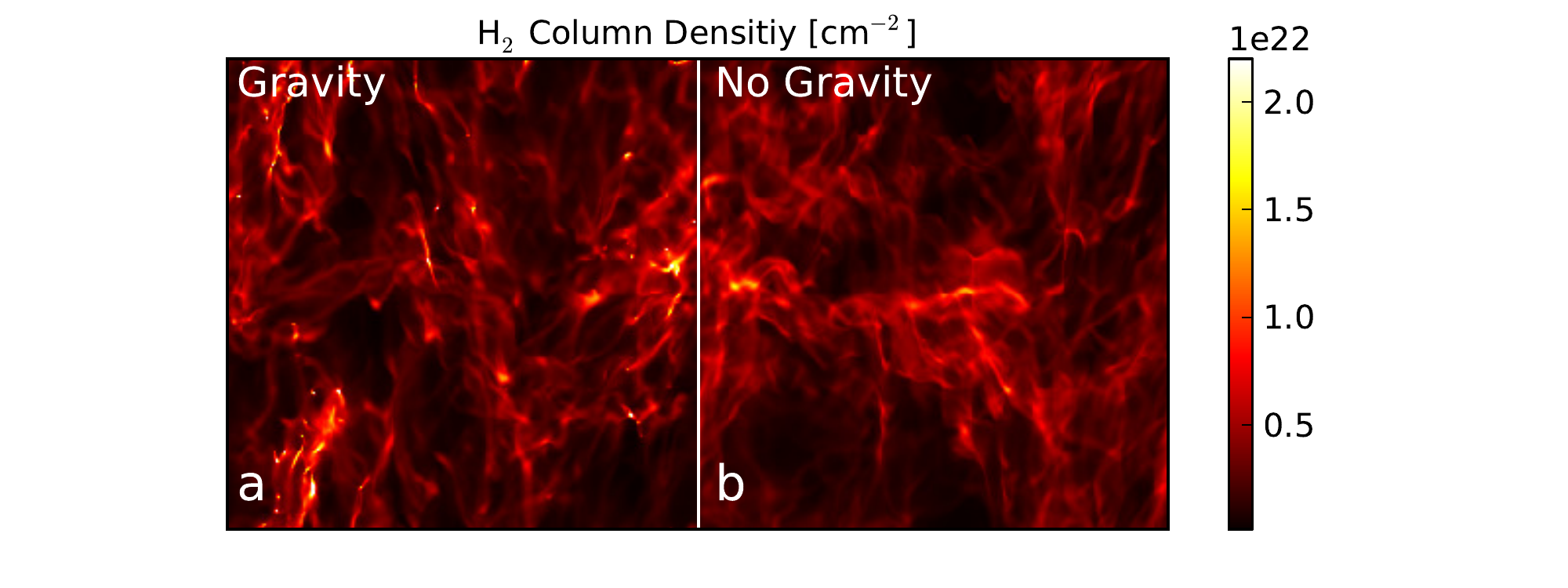}
\caption{The H$_2$ column density distribution in O1 at a) t=2.5 Myr, and b) t=0 Myr (the instant gravity is turned on)}
\label{fig:stella_moments}
\end{figure*}

The O1 simulation includes gravity. Gravity acts to gather and collapse gas, which may create locally crowded regions of high confusion. On the other hand, gravitational collapse should also gather diffuse material on large scales. This may decrease the filling factor on large scales, in mitigate confusion. To probe how important each of these effects are, Figure \ref{fig:virialvsvirial_nog} shows the equivalent set of comparisons of O1 at an earlier epoch, at the instant gravity is enabled. This timestamp captures the steady-state turbulent structure of the O1 simulation, before gravity has had an influence. The column density of the simulation at the original and earlier timestamp is shown in Figure \ref{fig:stella_moments}.

Gravity causes structure collapse at small scales, producing more dense, small regions (Figure \ref{fig:stella_moments}a). Before gravity is enabled, structures on average are larger, more diffuse, and overlap more. This effects the kinematic properties in Figure \ref{fig:virialvsvirial_nog} in the following ways: without gravity, there are fewer structures overall (147 vs 191 at the original simulation time). Because structures overlap more without gravity, there is a greater fraction of $q < 0.5$ structures (37\% vs 28\%). Finally, the virial parameter averaged over all structures is 5.8 without gravity, compared to 3.0 for the original O1 simulation.

Despite the moderately worse confusion, the scatter and bias in Figures \ref{fig:virialvsvirial} and \ref{fig:virialvsvirial_nog} are remarkably similar. In other words, the influence of gravity in the O1 simulation does not greatly impact the ability to recover physical properties.

\subsection{Effect of Chemistry}
A main difference between the O1 and S11 simulations is the inclusion of limited CO chemistry in S11. Spatial abundance variations in CO can decouple the CO density from the H$_2$ density. This, in turn, can break the correspondence between CO intensity and PPP density.

\begin{figure*}[htbp]
\includegraphics[width=3.4in]{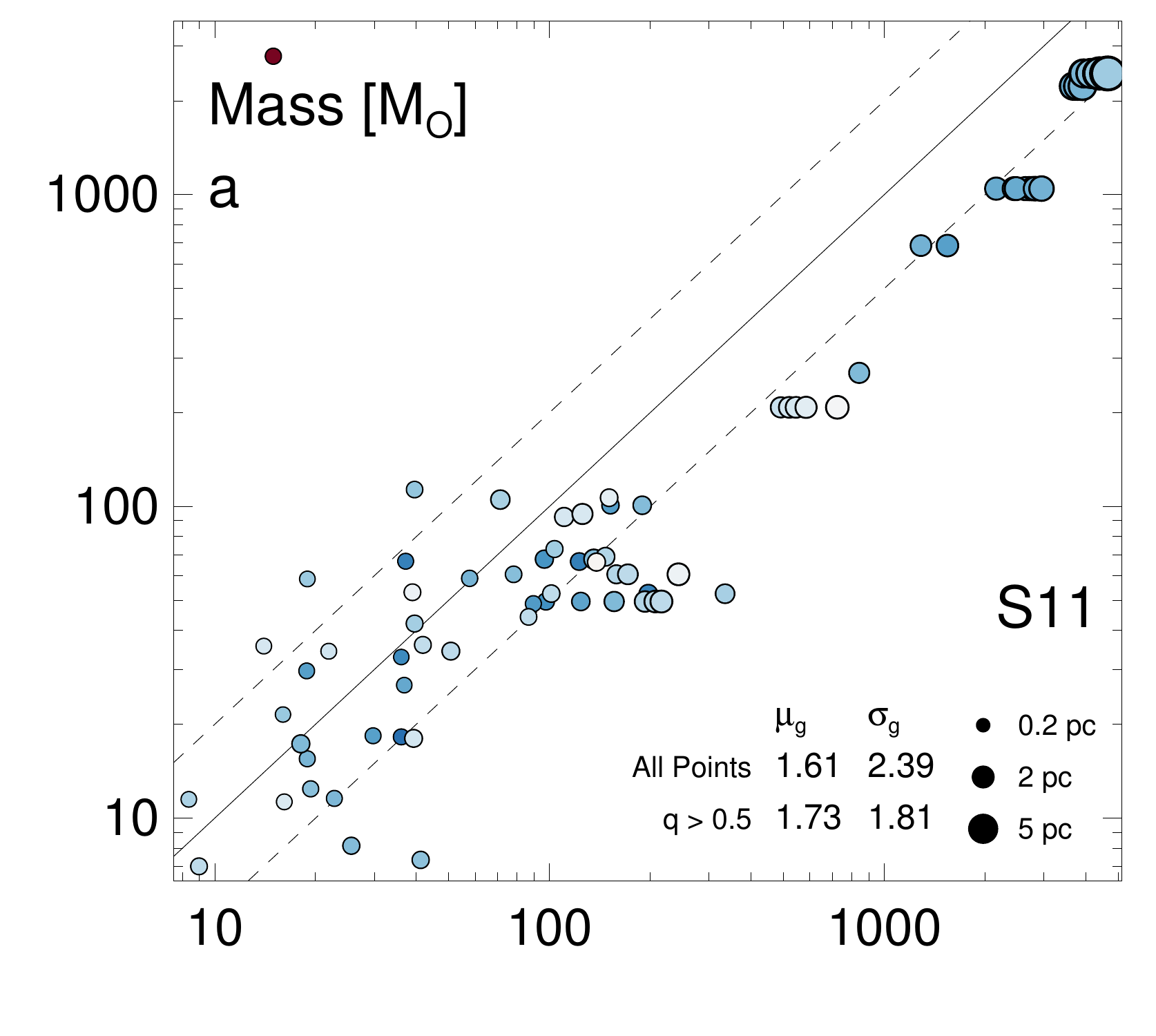}
\hfill
\includegraphics[width=3.4in]{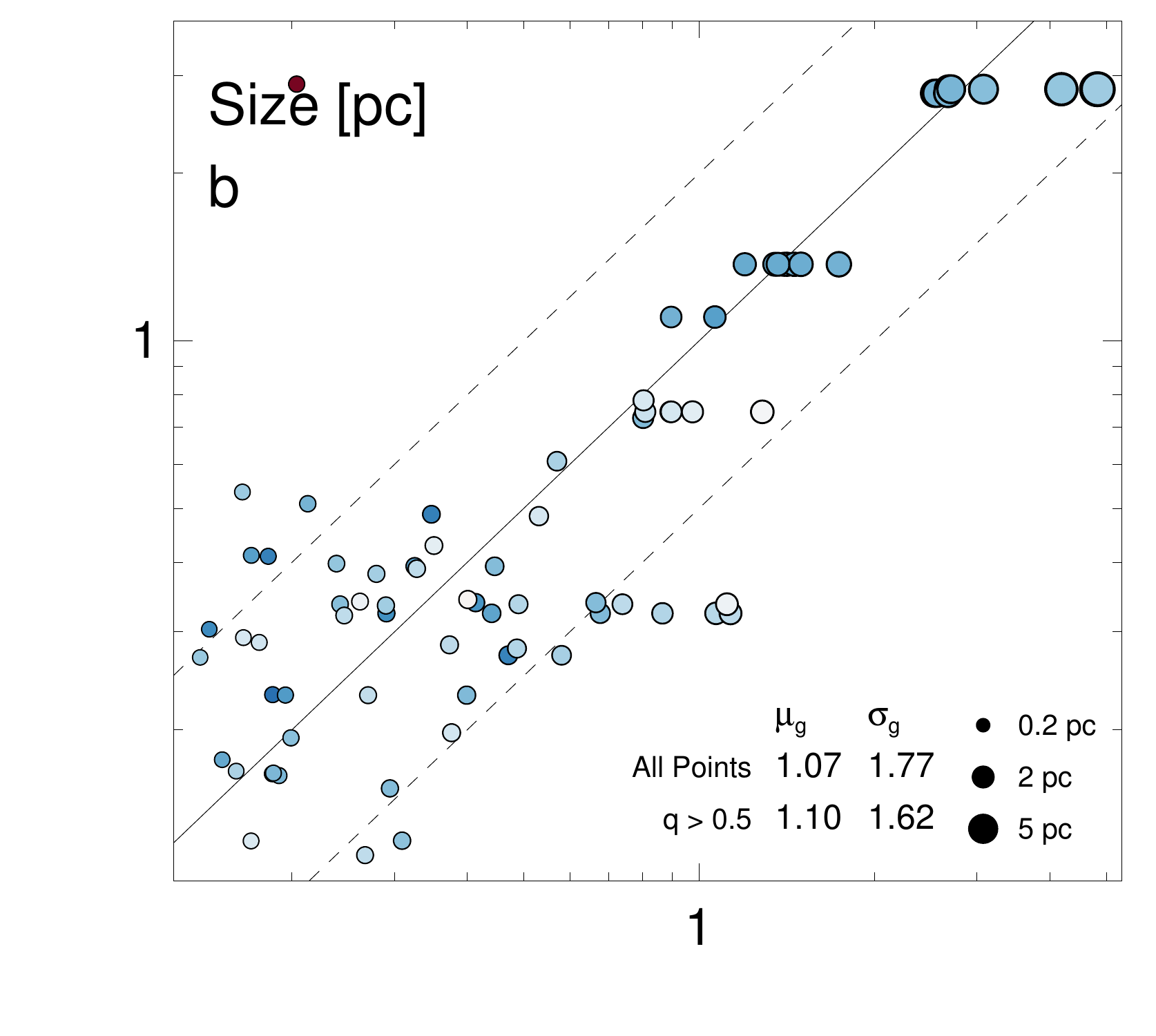}
\includegraphics[width=3.4in]{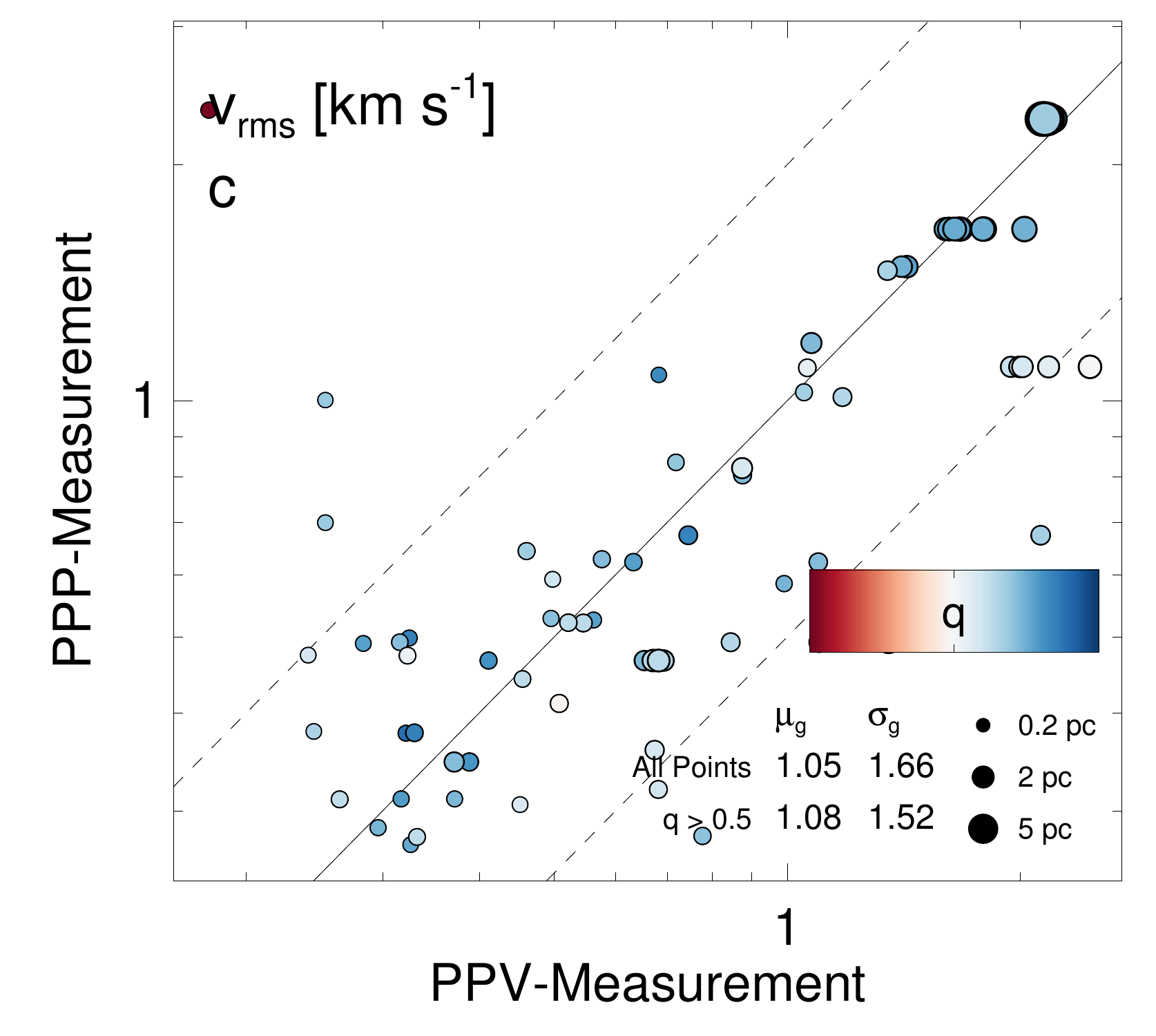}
\hfill
\includegraphics[width=3.4in]{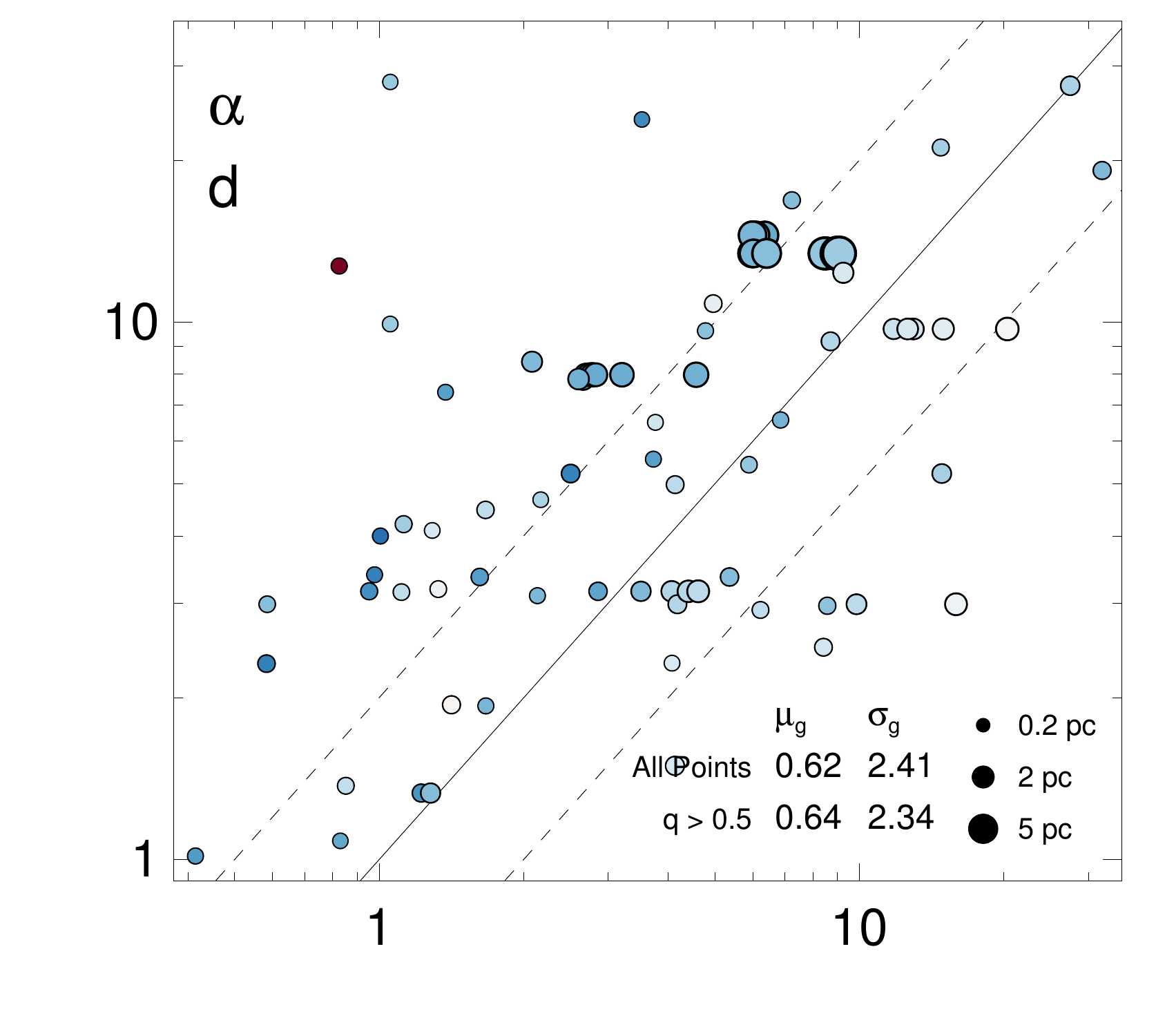}
\caption{The same as Figure \ref{fig:virialvsvirial}, for the \coc\, transition of the S11 simulation.}
\label{fig:virialvsvirial_rahul}
\end{figure*}

Figure \ref{fig:virialvsvirial_rahul} shows the same comparisons for the S11 simulation. Panels b and c have a comparable amount of bias and scatter as the previous figure for O1, albeit with fewer structures overall. The most dramatic difference between this plot and Figure \ref{fig:virialvsvirial} is the mass comparison in panel a. The points in the S11 simulation are shallower than the 1:1 line -- while PPV-structures recover sensible masses below $\sim 100 M_\odot$, $M_{\rm PPV}$ overestimates $M_{\rm PPP}$ for larger structures. This directly affects the virial plot in panel d, which exhibits a bias towards PPV-underestimates of $\alpha$.

One may also wonder if the approximation to estimate mass from intensity (given by Equation \ref{eq:xfactor_mass}) contributes to the mass discrepancy. This is unlikely. Masses are estimated by measuring the mass-to-light ratio of the brightest 5\% of the pixels and using this as a conversion factor. This conversion factor underestimates the mass-to-light ratio for low-column density lines of sight, where gas is sub-thermally excited and emits inefficiently. Adopting a more appropriate mass-to-light ratio for these lines of sight would actually lead to even larger $M_{\rm PPV}$ measurements, which exacerbates the discrepancy.

\begin{figure*}[htbp]
\includegraphics[width=7in]{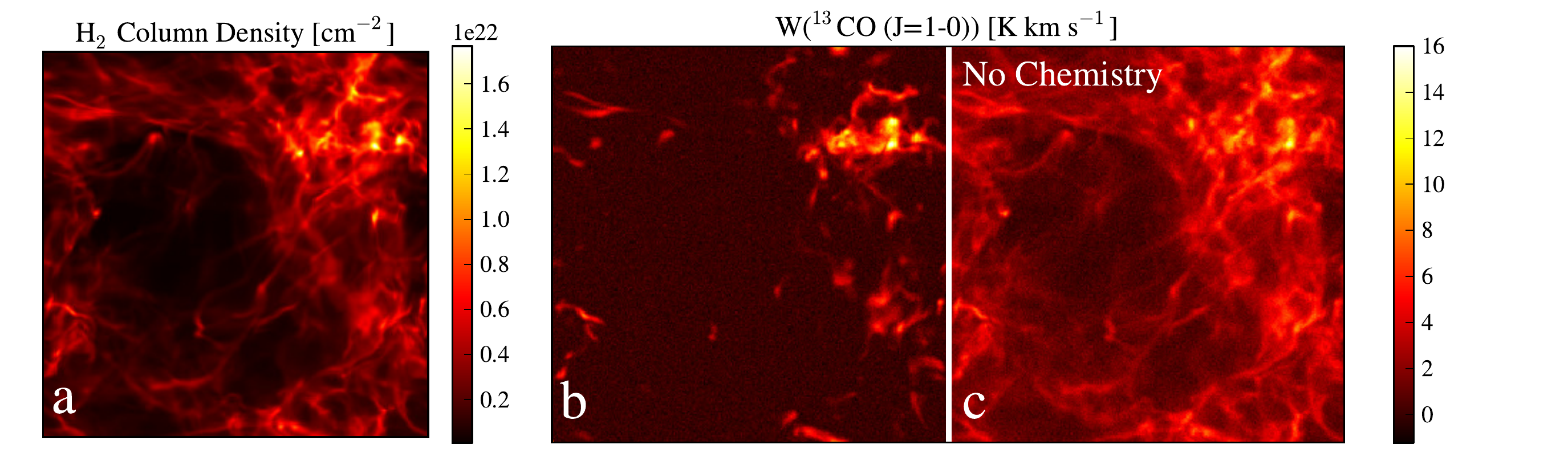}
\caption{The H$_2$ column density map of S11 (a), and the integrated $^{13}$CO (J=1-0) maps with and without chemistry (b, c).}
\label{fig:rahul_moment}
\end{figure*}

The mass discrepancy for large structures is caused by CO dissociation at low column densities, which S11 includes but O1 does not. Figure \ref{fig:rahul_moment}b shows that the S11 simulation contains large regions devoid of CO. There is H$_2$ in these regions (panel a), but little CO due to dissociation. In other words, the topologies of CO and H$_2$ gas partially diverge on large scales. It is thus more difficult to find exact PPP-equivalents of large-scale PPV structures in S11. This leads to discrepancies which, evidently, are most pronounced for mass measurements. This doesn't explain why the bias is towards PPV mass overestimates, as opposed to underestimates. We do not have a simple explanation for the direction of the bias. For whatever reason, PPV structures are better matched (as quantified by Equation \ref{eq:match}) to slightly too-small PPP structures than they are to slightly too-large PPP structures. We speculate this has to do with the detailed topology of gas in PPP vs PPV.

\begin{figure*}[htbp]
\includegraphics[width=3.4in]{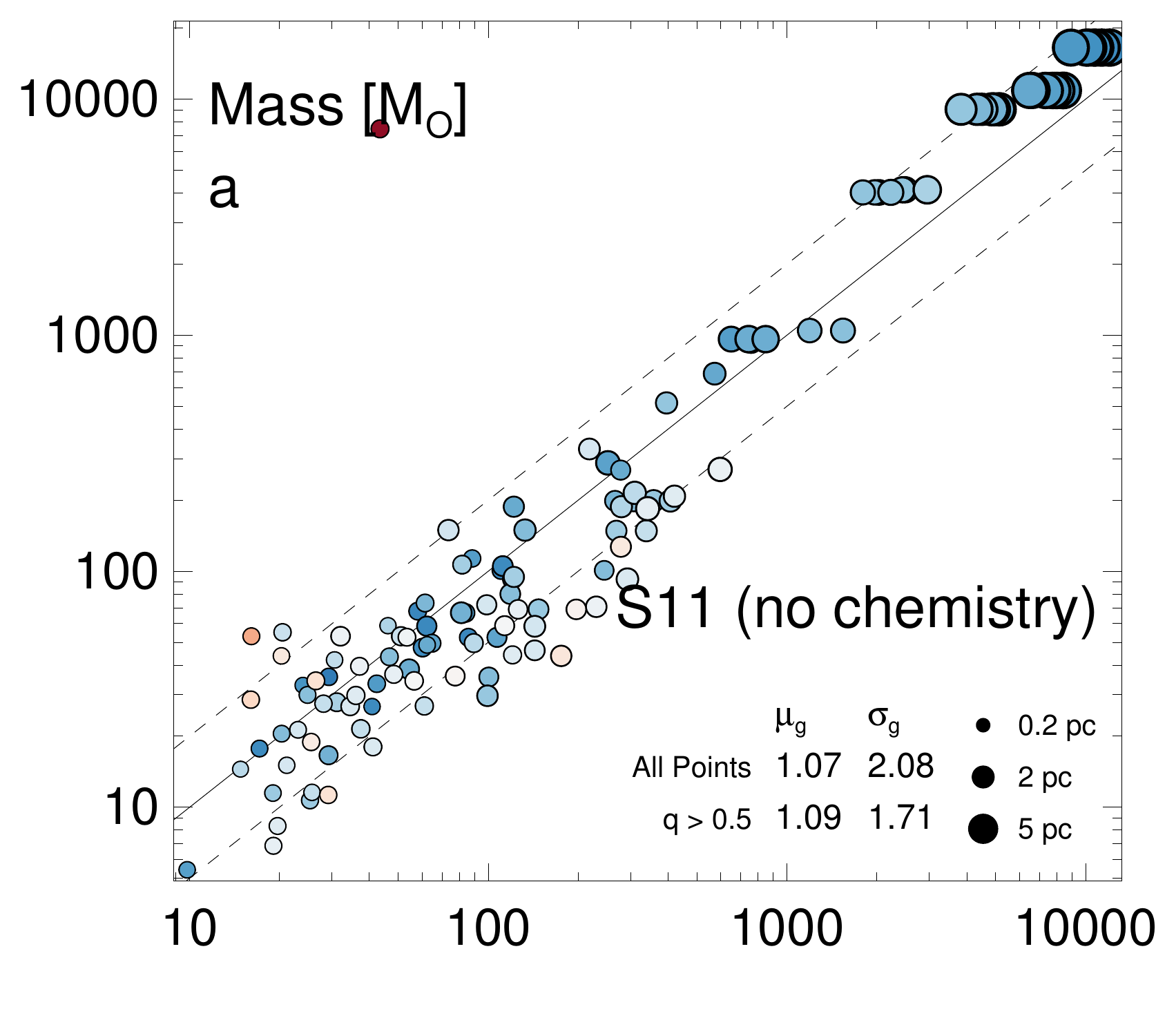}
\hfill
\includegraphics[width=3.4in]{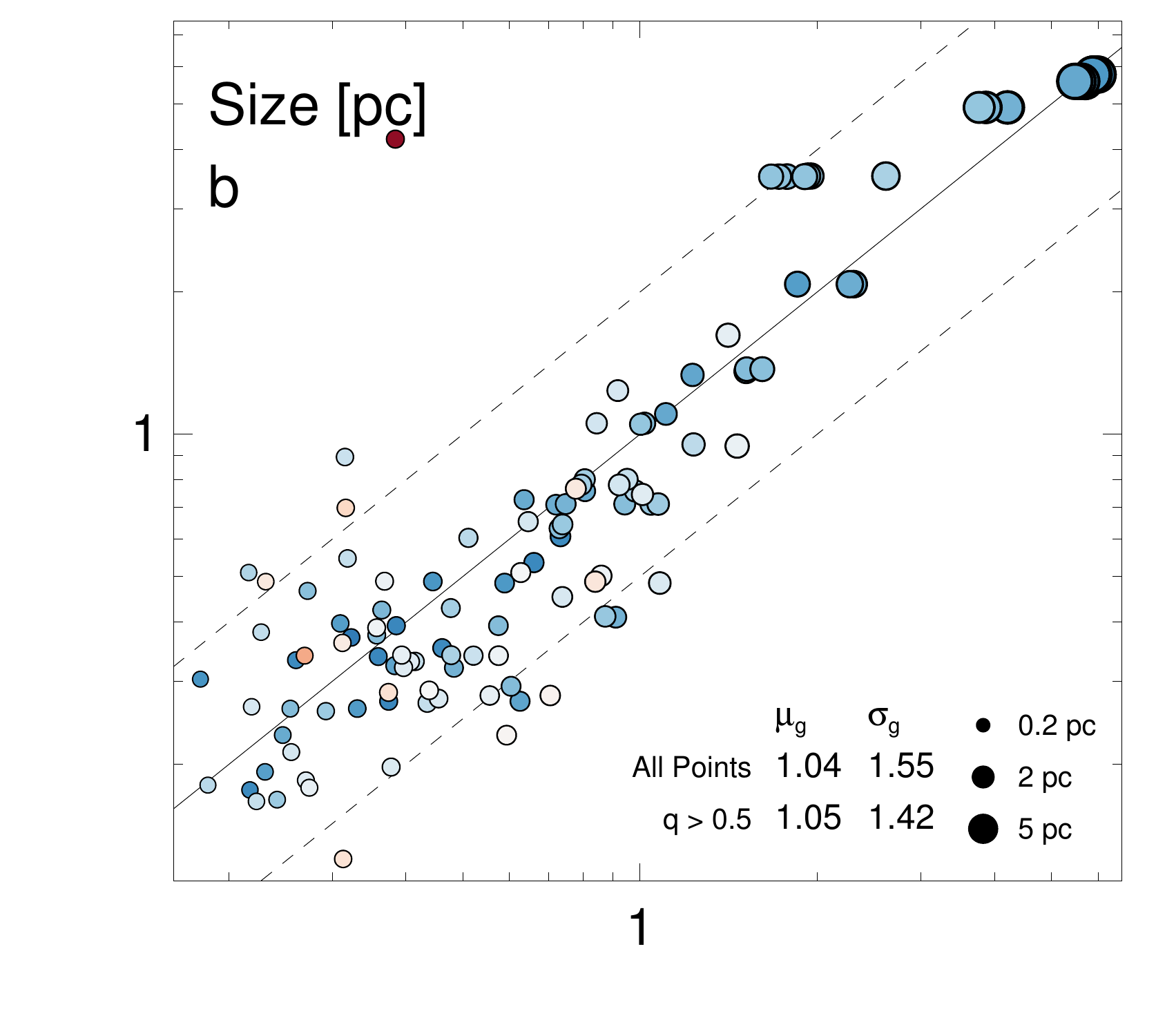}
\includegraphics[width=3.4in]{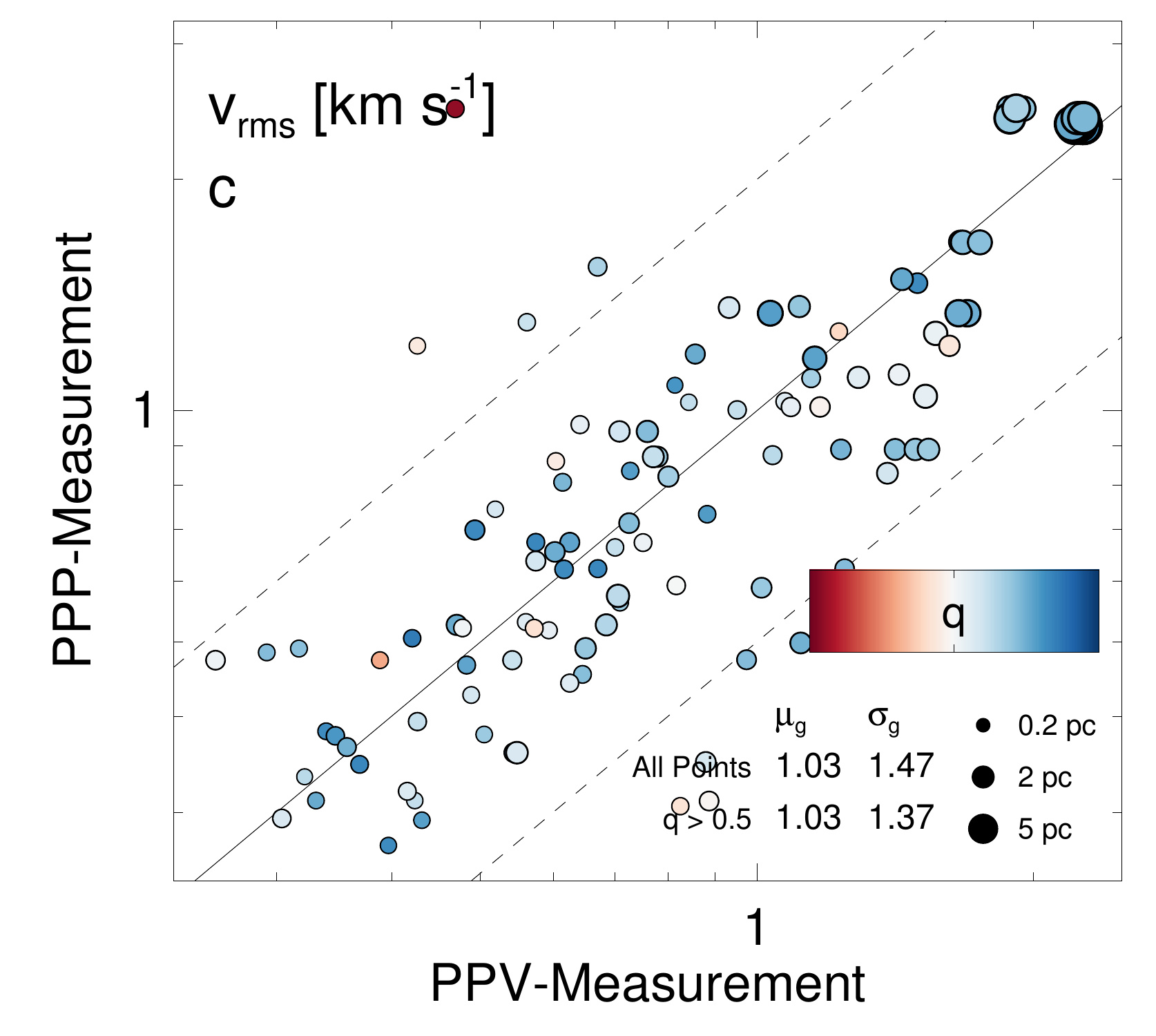}
\hfill
\includegraphics[width=3.4in]{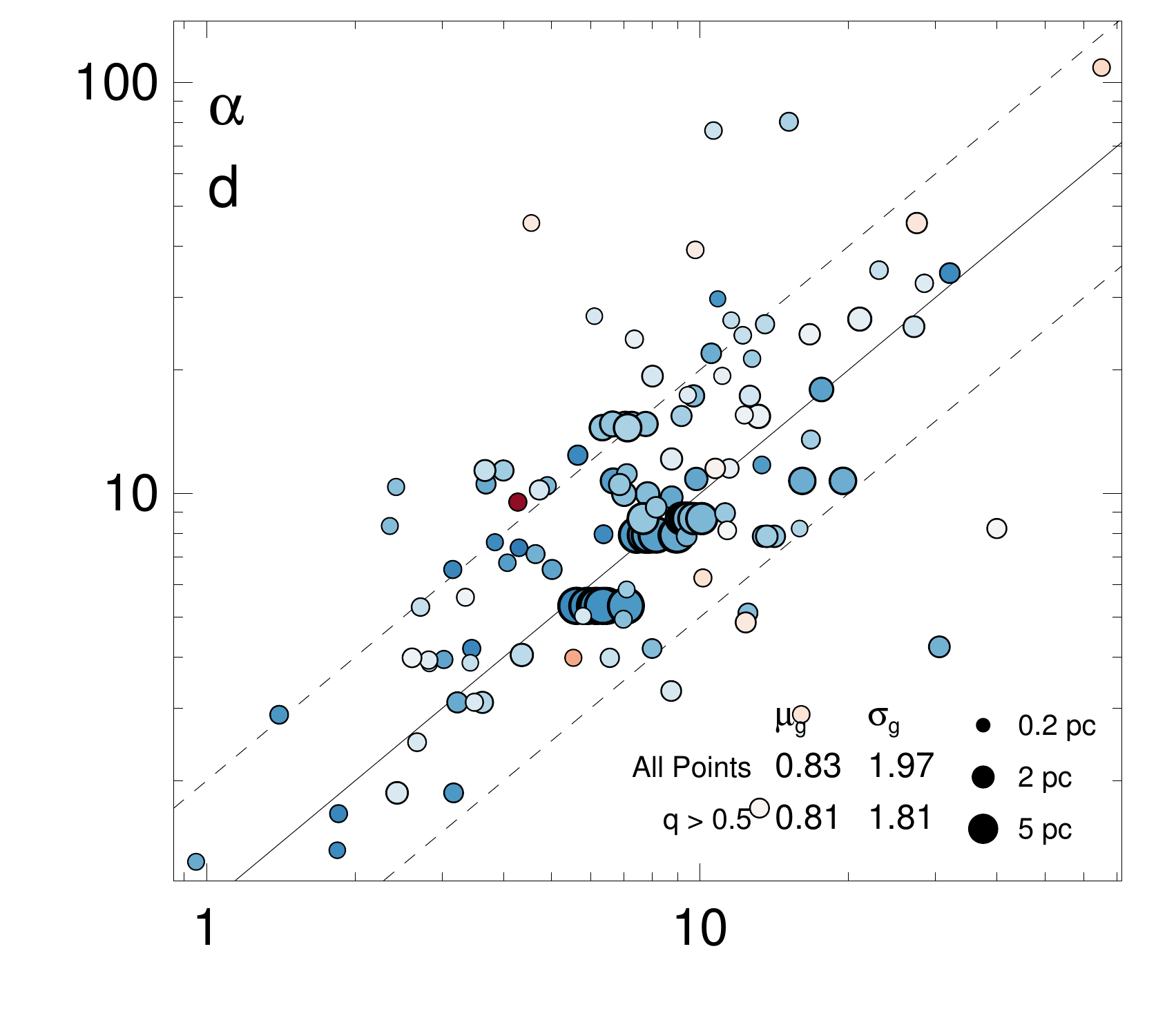}
\caption{The same as Figure \ref{fig:virialvsvirial_rahul}, but for the S11 simulation with no chemistry.}
\label{fig:virialvsvirial_nochem}
\end{figure*}

We can verify that chemistry in S11 causes the mass discrepancy by repeating the analysis on a version of S11 without chemistry. To do this, we re-compute the radiative transfer on S11, assuming a constant temperature of 15K, and a constant CO/H$_2$ abundance of 10$^{-4}$. The resulting integrated intensity map is shown in Figure \ref{fig:rahul_moment}c, and the kinematic comparisons in Figure \ref{fig:virialvsvirial_nochem}. There are more structures in this simulation due to extra CO emission. Furthermore, the mass-mass plot follows the 1:1 line much better.

At the largest scales, the S11 simulation (which does not include gravity) has a virial parameter of $\alpha_{\rm PPV} \sim 9$. This is higher than the O1 simulation, for which the largest-scale structures have a virial parameter of $\alpha_{\rm PPV} \sim 1$. The S11 structures with the smallest values of $\alpha_{\rm PPV} \sim 1$ are in fact cause for concern, since they indicate that gravitational and kinetic energies are comparable in magnitude. Because S11 does not include the effects of gravity, the dynamical nature of these structures is less faithfully modeled.

\subsection{Assessing Boundedness}

The virial parameter is most often used to estimate the gravitational boundedness of a structure, with values of $\alpha < 2$ interpreted to indicate that a structure is bound. This interpretation is problematic, as it ignores other forces and oversimplifies the role of turbulence as a support against collapse \citep{http://adsabs.harvard.edu/abs/1992ApJ...395..140B, http://adsabs.harvard.edu/abs/2006MNRAS.372..443B}. Our kinematic analysis of S11 and O1 add another cause for concern: measurements of the virial parameter based on PPP and PPV data differ by a factor of $2-3$, and cluster around $\alpha_{\rm PPV} = 1-5$. This implies that, using CO data alone, it is unclear on which side of the $\alpha_{\rm PPP} = 2$ boundary many structures fall.

To illustrate this, we repeat an analysis similar to \cite{http://adsabs.harvard.edu/abs/2009Natur.457...63G}, who 
measured the fraction of low-virial parameter structures for the L1448 subregion of Perseus as a function of size.

For each simulation, we assign a virial parameter to each voxel according to the smallest structure to which that voxel belongs. Next, we bin the structures by size, and in each bin, make a mask of all the pixels associated with these structures; we refer to this set as $\mathcal{S}$. Finally, we compute the fraction of emission in these pixels with $\alpha_{\rm PPV} < 2$:

\begin{equation}
f = \frac{\sum{\{L(\vec{r}) | \vec{r} \in \mathcal{S} , \alpha(\vec{r}) < 2}\}}{\sum{\{L(\vec{r}) | \vec{r} \in \mathcal{S}}\}}
\label{eq:frac}
\end{equation}

\begin{figure}[htbp]
\centering
\includegraphics[width=3.5in , trim=0 .5in 0 0, clip]{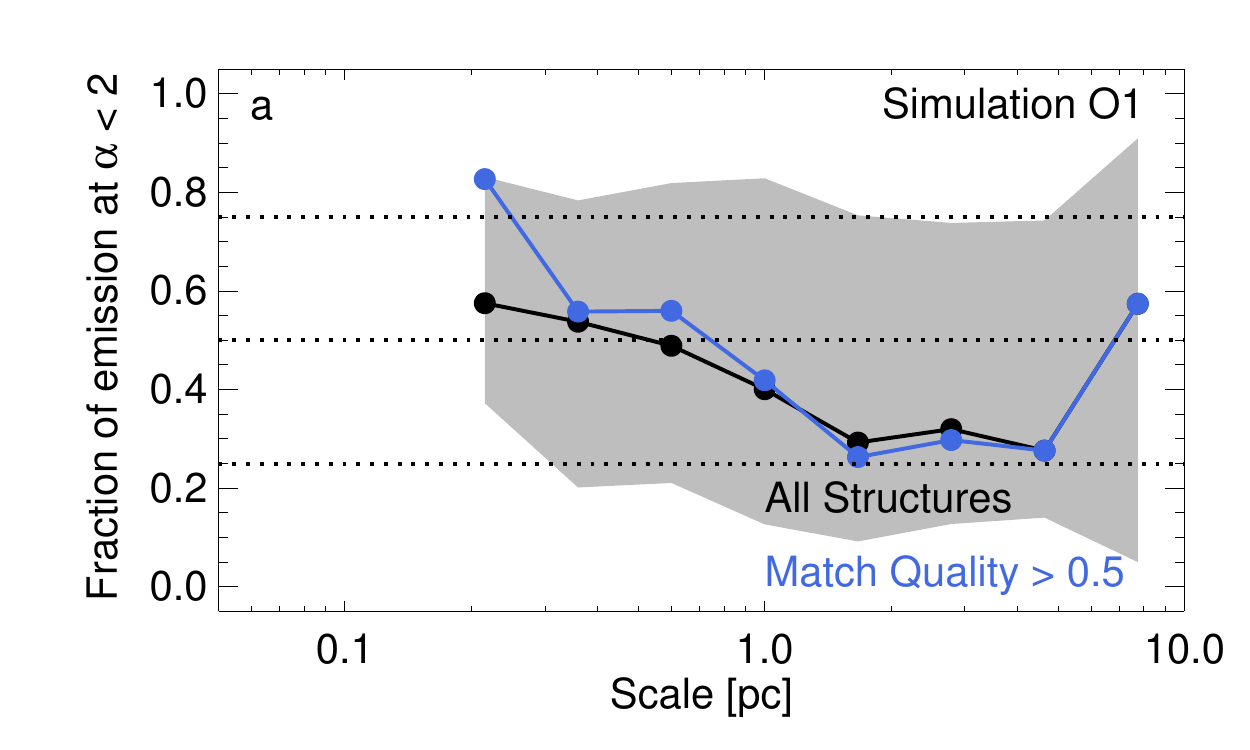}
\includegraphics[width=3.5in,  trim=0 .5in 0 0, clip]{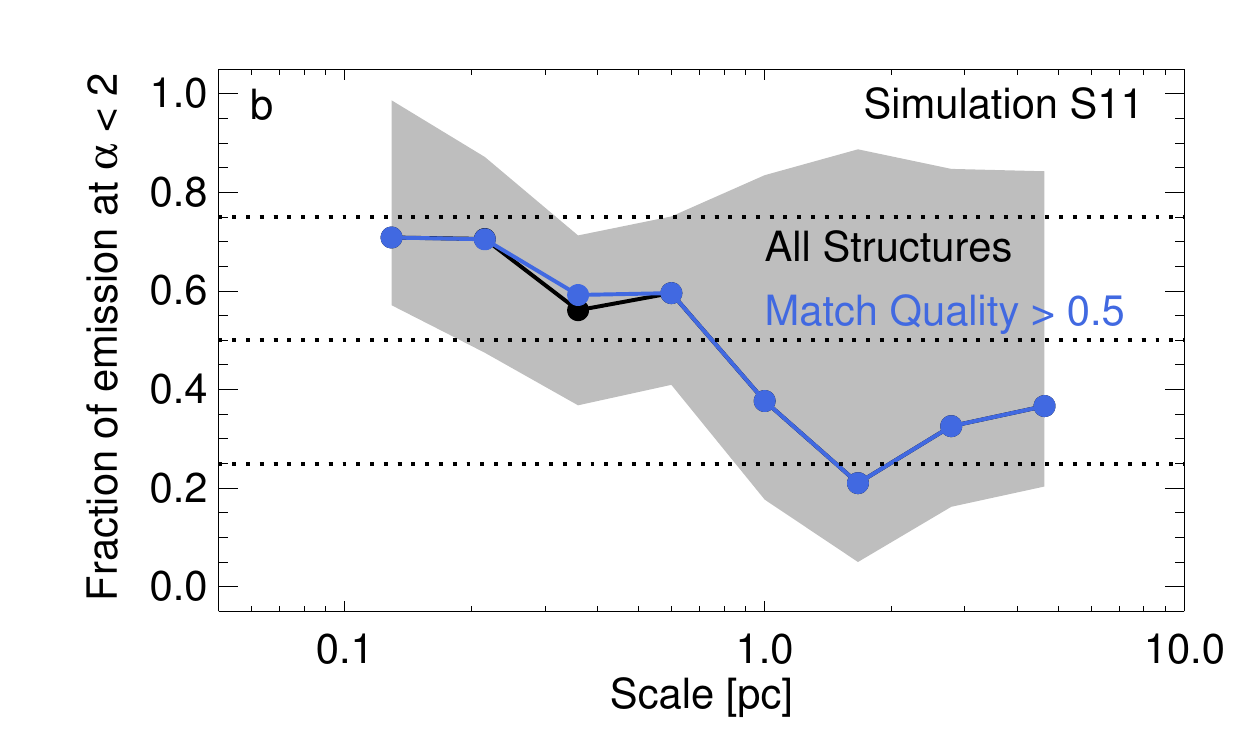}
\includegraphics[width=3.5in,  trim=0 .5in 0 0, clip]{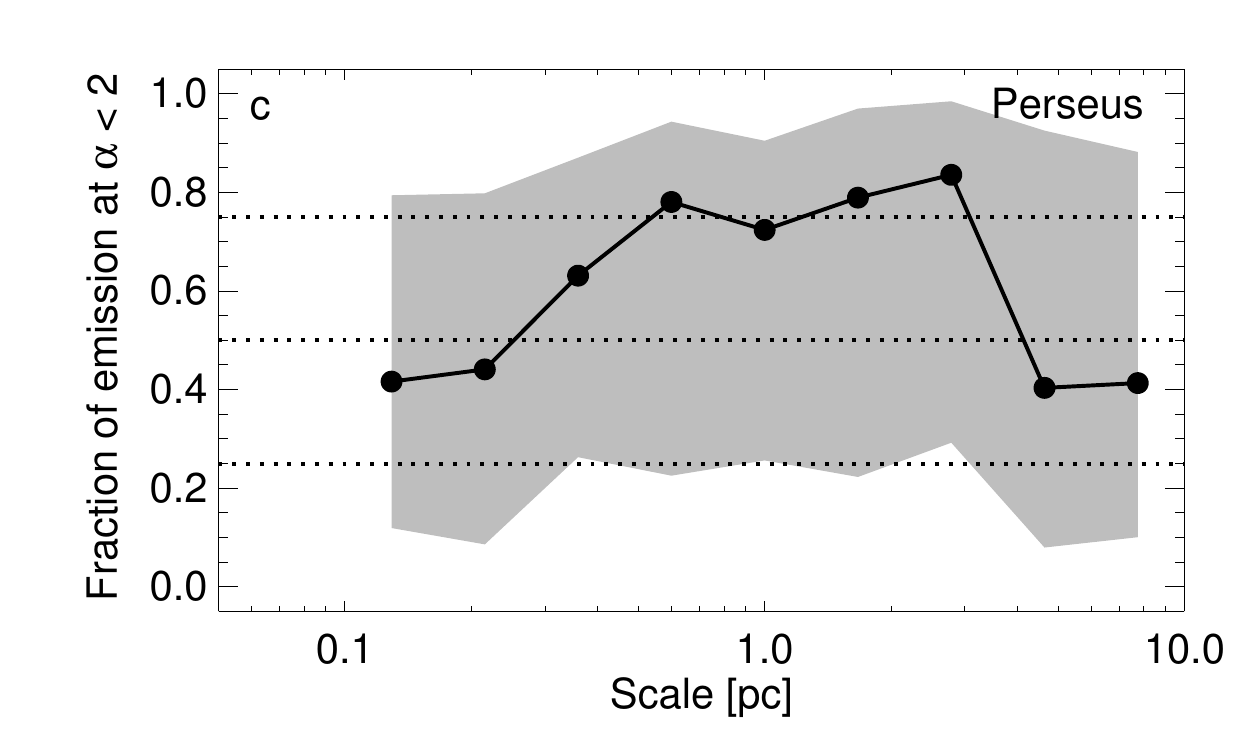}
\includegraphics[width=3.5in]{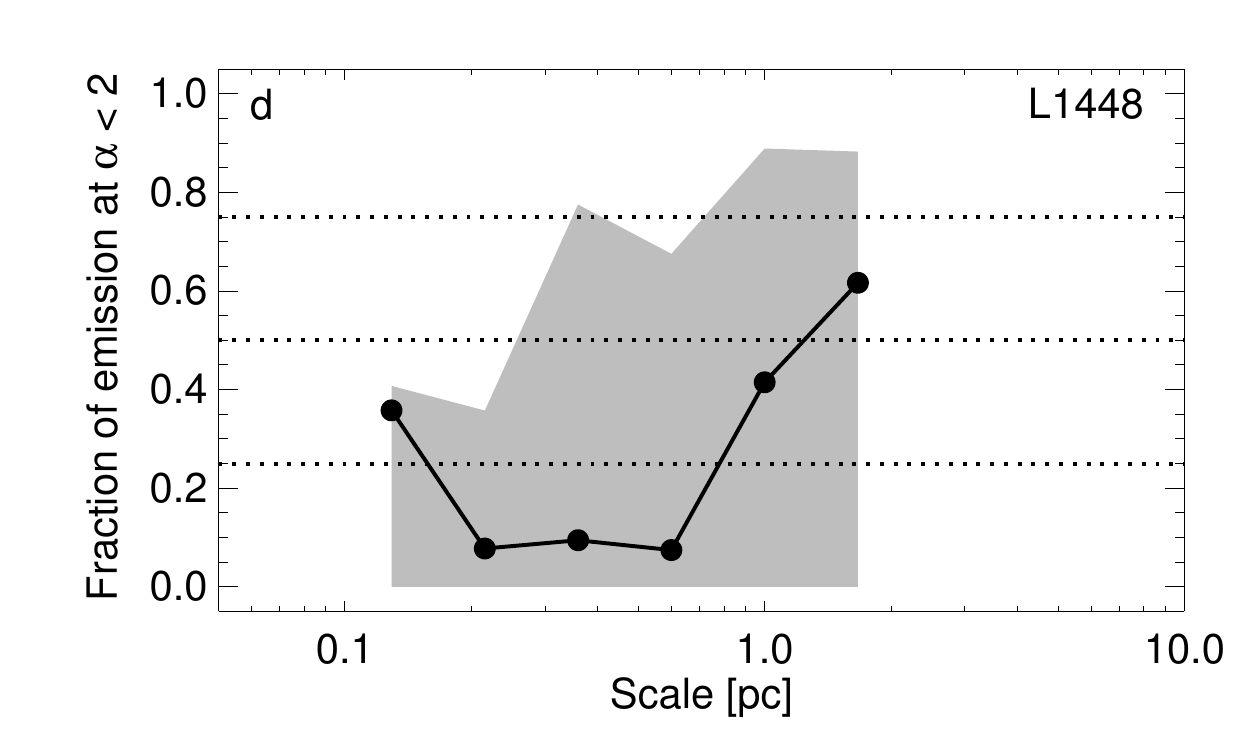}

\caption{The fraction of emission in \(\alpha_{\rm PPV} < 2\) structures, as a function of structure size, derived from \coc. The four panels show the O1 simulation, S11 simulation, Perseus as a whole, and L1448.  For the simulations, the blue line traces the relationship for structures with match qualities \textgreater 0.5.}
\label{fig:size_virial}
\end{figure}

We plot this fraction as a function of size for O1, S11, Perseus, and L1448 in Figure \ref{fig:size_virial}. Note that the L1448 plot is slightly different from Figure 4 in \cite{http://adsabs.harvard.edu/abs/2009Natur.457...63G} because we use a different scheme
for measuring the fraction of $\alpha_{\rm PPV} < 2$ emission\footnote{Namely, Goodman et al. sum over structures, whereas we sum over pixels.}. Also, remember that $\alpha_{\rm PPV}$ underestimates $\alpha_{\rm PPP}$ in the S11 simulation, pushing the line higher than it would otherwise be.

For the O1 and S11 simulations in Figure \ref{fig:size_virial}a and b, we plot the relationship for all structures (black), as well as
those with high match qualities ($q > 0.5$, blue). Because estimates of $\alpha$ in PPV are scattered by a factor of $\sim2$ about the
corresponding PPP measurements, the grey bands show the range of possible values the black line can take if each value of $\alpha$ is mis-estimated by a factor of 2. Because so many structures fall within a factor of 2 of $\alpha_{\rm PPV}=2$, the grey band covers a large swath of the plot. Thus, in addition to the conceptual problems associated with inferring boundedness from the value of $\alpha_{\rm PPV}$, \textit{there is an intrinsic observational ambiguity associated with reliably determining structures to be above or below $\alpha=2$.}

\subsection{Generalizing to Real Data}
\label{sec:generalize}

As discussed in Section \ref{sec:perseus}, the O1 and S11 simulations
do not reproduce several statistical properties of Perseus. In
particular, both simulations tend to have lower line-of-sight
velocity dispersions than Perseus. This may act to increase
superposition effects in the simulations since the material is more
crowded in PPV space. Similarly, the lack of external radiation fields in O1 (with no chemistry) suppresses dissociation of low-density CO, producing an artificially high filling factor of emission and greater likelihood for superposition. Clouds with significant excitation
variation due to external heating may have less PPV superposition of features. 

As the field of molecular cloud simulation and synthetic observation advances, we should be able to make new quantitative estimates of the degree to which various tracers are confused under different physical and observing conditions. The suite of diagnostics presented in Appendix A enables a standardized comparison between a simulated and observed dataset. These comparisons address the main statistical cloud properties most relevant for superposition analysis -- column density, intensity, and linewidth -- and can help to assess how well a given simulation acts as a surrogate for studying unobservable projection effects in a real dataset.

\section{Conclusion}
Intensity features in molecular cloud observations do not always correspond neatly to real density structures. The degree of correspondence is difficult to assess observationally or characterize analytically, but it can be measured in simulations. Such a comparison helps develop intuition about problems when interpreting real datasets. 

We have conducted such a comparative analysis in this paper, presenting a new technique to cross-match PPP density structures with PPV intensity structures in synthetic molecular cloud observations. This gives a structure-by-structure assessment of how well cloud properties are recovered in observations. In particular, we find that:

\begin{enumerate}
\item {Structures traced in CO are more distorted in observations of more space-filling emission, so that the \coa\, transition shows the most severe effects of overlap, while \cob\, is less affected, and \coc\, gives the most faithful representation of PPP structures in PPV space. This is primarily due to the opacity of the lines, which obscures density structures in the back of the cloud.}
\item Comparing size, mass, velocity dispersion, and virial parameter as measured in PPP (real) and PPV (observed) space, we find that size, mass, and velocity dispersion can usually be recovered to within 40\%. Measurements of the virial parameter have a larger scatter of 0.3 dex (a factor of 2).
\item The uncertainty in recovering the virial parameter from CO measurements imposes an unavoidable ambiguity about the energy balance of many cloud structures. In particular, it is often ambiguous to which side of the $\alpha = 2$ threshold most cloud substructures fall. Thus, assessing the relative dominance of gravitational versus kinetic energy is difficult, as is assessing boundedness.
\item In the simulations studied here, most molecular cloud structures have PPV-measured virial parameters within a factor of 2 of $\alpha_{\rm PPV} = 2$. Thus, if projection effects induce a factor of 2 uncertainty on $\alpha$, there is a large ambiguity regarding which substructures in a cloud are ``bound'' in the sense that $\alpha < 2$.
\item Gravity can act to modestly reduce confusion, by gathering material into more compact, less-overlapping structures. However, this does not have a significant impact on the precision to which intensity structures can be recovered from CO measurements.
\item The primary impact of chemistry is lower the abundance of CO at low column densities and create excitation temperature variations. This reduces the optical depth and, hence, reduces the amount of confusion, but it may also decouples the topology of CO emission from the underlying H$_2$ density. This most heavily affects structures with $M \gtrsim 100M_\odot$, leading to factors of 2-3 discrepancies between PPV-derived and PPP-derived masses.
\end{enumerate}

Simulations can be powerful probes of otherwise-unobservable phenomena that affect real data. However, conclusions drawn from such analyses are
limited by how well simulations approximate the observed properties of real clouds. The simulations in this work do not reproduce several of the details of the emission properties of Perseus (in particular the characteristic brightness and velocity dispersion of CO lines). We conclude that simulations which initially appear to be qualitatively similar to an observed cloud can be surprisingly different in detail. The direct comparison of simulations and observations is fraught with subtleties, and much care must be taken to obtain true quantitative agreement. We advocate for future studies to examine PPP-PPV issues in more detail, including the production of simulations that are more representative of well-characterized molecular clouds like Perseus.

We thank  Jens Kauffmann, Lukas Konstandin, Ralf Klessen, Eve Ostriker,  Erik Rosolowsky, and Mark Heyer (the referee), whose comments improved this manuscript. Support for this work was provided by NASA through Hubble Fellowship grant \#HF-51311.01 awarded by the Space Telescope Science Institute, which is operated by the Association of Universities for Research in Astronomy, INC., for NASA, under contract NAS 5-26555 (SSRO). RS and SG acknowledge financial support from the Deutsche Forschungsgemeinschaft (DFG) via SFB 881 ``The Milky Way System'' (sub-projects B1 and B2).

\input{bib.tex}
\appendix{

\section{Diagnostics}
\label{sec:diagnostics}

This paper uses simulations as surrogates for molecular clouds like
Perseus, to better understand how unobservable effects like
superposition affect subsequent analyses. These results generalize to
real data only to the extent that the simulations share the same
observable properties as real clouds. These simulations are approximate analogs to molecular clouds like Perseus. 
However, they still show several discrepancies with Perseus when examined in
detail. We present several diagnostics in this appendix, in part to define
a standard set of criteria that can be used to evaluate the
observational applicability of cloud simulations. We also
propose a way to reduce each diagnostic to a single score from 0-1, to
more easily communicate how well a given simulation reproduces a particular
property of a real cloud observation.

Figures \ref{fig:dgrid_stella} and \ref{fig:dgrid_rahul} summarize the diagnostic comparisons for the O1 and S11 simulations.

\subsection{Data Filtering}
The metrics that follow measure properties along lines-of-sight. As such, we try to focus only on lines-of sight with substantial cloud emission and mask out
regions with little cloud material. We base our masking on the process described in \cite{http://adsabs.harvard.edu/abs/2008ApJ...679..481P} and require
each line of sight to satisfy the following inequalities:

\begin{enumerate}
\item The peak  \coa\, line intensity is at least 10$\sigma$ above the T=0.
\item The peak \coc\, line intensity is at least 5$\sigma$ above T=0.
\item The velocity dispersion of \coa\, is at least 0.8 $\times$ the velocity dispersion of \coc.
\end{enumerate}

These cuts are applied both to the Perseus data and to each simulation. The first two cuts are self explanatory. Pineda et al. proposed the last cut to further filter noisy or pathological lines of sight; the rationale is that, since $^{12}$CO is more abundant and opaque than $^{13}$CO, it should always have a larger spatial and kinematic extent.

\subsection{Column Density Distribution}
The distribution of column densities is reasonably well-characterized for nearby clouds: near-infrared extinction measurements trace column densities
across the range $10^{20} \lesssim N_{\rm H_2}~  [{\rm cm}^{-2}]  \lesssim 10^{23} $, and far infrared dust emission probes higher column densities \citep{http://adsabs.harvard.edu/abs/2009ApJ...692...91G, http://adsabs.harvard.edu/abs/2012ApJ...752...55K, 2009A&A...493..735L, 2009A&A...508L..35K}. 

Most cloud column density distributions are approximately log-normal, with mean column densities of $N_{\rm H_2} \sim 10^{21}$ cm$^{-2}$ and width parameters $\sigma \sim 0.3-0.5$. In addition, some clouds (especially those currently undergoing star formation) display excess power-law tails at column densities above $\sim 3 \times 10^{21} {\rm cm}^{-2}$  \citep{http://adsabs.harvard.edu/abs/2009ApJ...692...91G, 2009A&A...508L..35K}.

The column density distribution is shown in Figure \ref{fig:dgrid_stella}a for O1, and \ref{fig:dgrid_rahul}a for S11.

\subsection{Integrated Intensity Distribution}
The distribution of line-of-sight integrated intensity $W=\int I(x, y, v)\, {\rm d}v$  is also straightforward to compute from spectral line observations. Its distribution is shown in Figure  \ref{fig:dgrid_stella}b and \ref{fig:dgrid_rahul}b for \coa, and  Figure \ref{fig:dgrid_stella}c and \ref{fig:dgrid_rahul}c for \coc.

\subsection{Distribution of Peak Intensity}
The distribution of peak line-of-sight intensity is a crude measure of the excitation state of the gas; it breaks the degeneracy between excitation state and column density in the integrated line intensity. Its distribution is shown in Figure  \ref{fig:dgrid_stella}d and \ref{fig:dgrid_rahul}d for \coa, and  Figure \ref{fig:dgrid_stella}e and \ref{fig:dgrid_rahul}e for \coc.
 
\subsection{Velocity Dispersion Distribution}
Likewise, we can compute the distribution of line-of-sight velocity dispersions. We compute the velocity dispersion by computing the intensity-weighted second moment of velocity along each line of sight. Since the second moment is sensitive to faint emission at large velocity offsets, we only consider pixels 3$\sigma$ above the background. Its distribution is shown in Figure  \ref{fig:dgrid_stella}f and \ref{fig:dgrid_rahul}f for \coa, and  Figure \ref{fig:dgrid_stella}g and \ref{fig:dgrid_rahul}g for \coc.

\subsection{Joint Distribution of Column Density and Line Intensity}
The previous diagnostics are 1-dimensional distributions, and say nothing about the correlation among different quantities. The joint distribution of line intensity and column density is particularly interesting, since the ratio of these quantities defines the much-studied X-factor. Higher line intensities \textit{at a given column density} indicate higher excitation levels, lower opacity, greater abundance of the exciting molecule, and/or greater linewidth (if the line is opaque). The joint distributions are shown in Figure  \ref{fig:dgrid_stella}h and \ref{fig:dgrid_rahul}h for \coa, and  Figure \ref{fig:dgrid_stella}i and \ref{fig:dgrid_rahul}i for \coc.

\subsection{Scoring Diagnostics}
Each of the above diagnostics can be converted into a numerical score, to quickly summarize how well a given simulation reproduces a given diagnostic. We base our score on the Kuiper statistic for two cumulative distribution functions:

\begin{equation}
K = \max {\rm \left(CDF_A - CDF_B \right) + \max \left( CDF_B - CDF_A \right) }
\end{equation}

The Kuiper statistic is a modification of the well-known Kolmogorov-Smirnov statistic, and is more sensitive to discrepancies in the tails of distributions (see the discussion in Section 14.3.4 of \citealt{Press07}).

For every comparison of 1-dimensional distributions, we define the score as $1 - K$, where smaller scores indicate less similarity between the simulation and Perseus.

There are a few ways to generalize the Kuiper statistic to the 2-dimensional joint distribution of column density and line intensity (see Section 14.8 of \citealt{Press07}). For each of these 2-dimensional distributions, we compute 4 cumulative distribution functions

\begin{eqnarray}
CDF_1(X, Y) &=& P(x < X, y < Y) \\
CDF_2(X, Y) &=& P(x > X, y < Y) \\
CDF_3(X, Y) &= &P(x < X, y > Y) \\
CDF_4(X, Y) &=& P(x > X, y > Y),
\end{eqnarray} 

compute the $K$ statistic for each CDF and save the largest statistic. Our final score is defined as $1 - {\rm K}_{max} / 2$. The factor of 2 correction is included because, in  two dimensions, the Kuiper statistic varies between 0--2 whereas, in 1 dimension, it varies between 0--1.

Table \ref{tab:scores} summarizes these scores for the O1 and S11 simulation, using Perseus as the benchmark. We encourage other researchers to generate molecular cloud simulations that better reproduce these observational diagnostics.

\input{diagnostic_table.tex}

\begin{figure}
\includegraphics[width=6in]{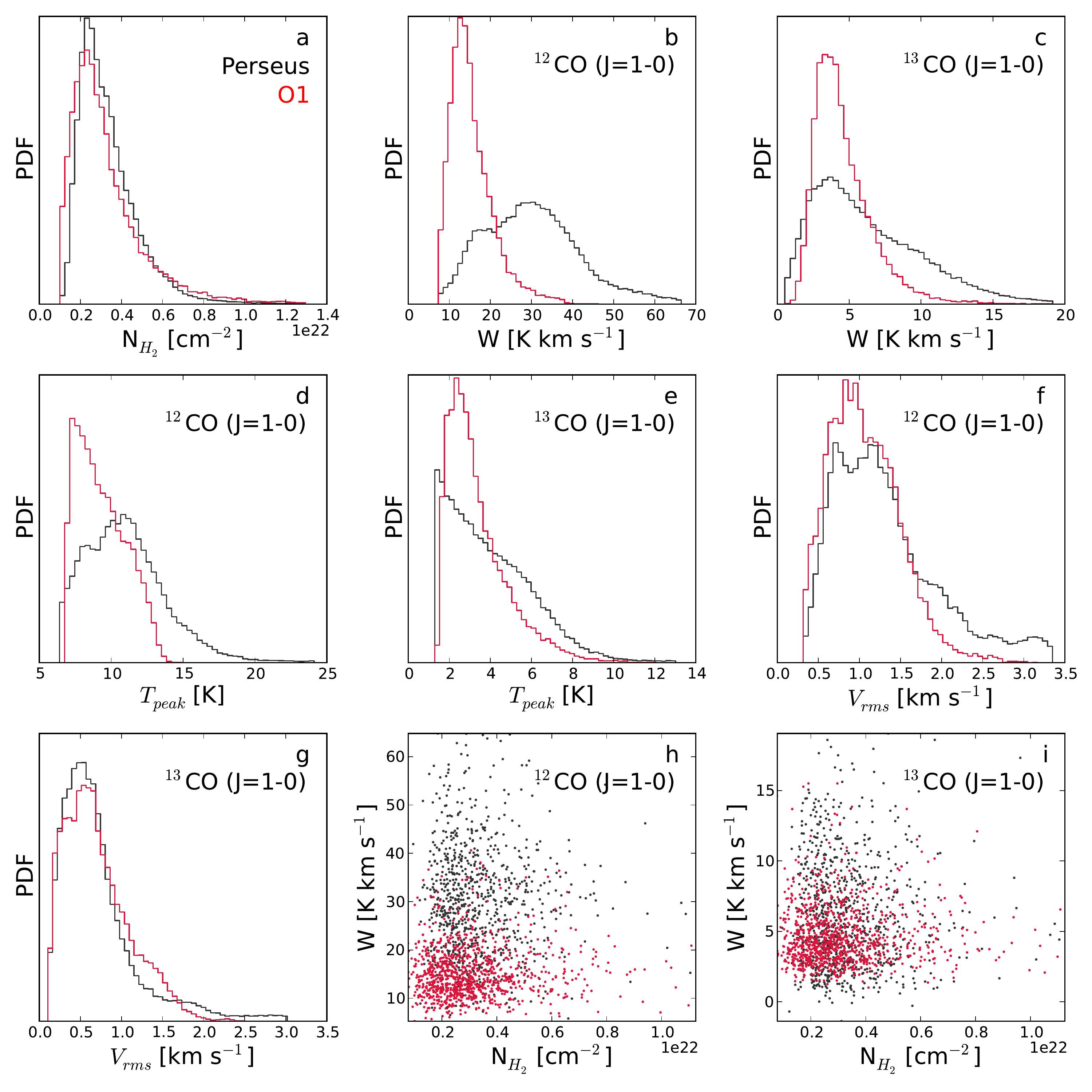}
\caption{The grid of diagnostic comparisons for the O1 simulation.}
\label{fig:dgrid_stella}
\end{figure}

\begin{figure}
\includegraphics[width=6in]{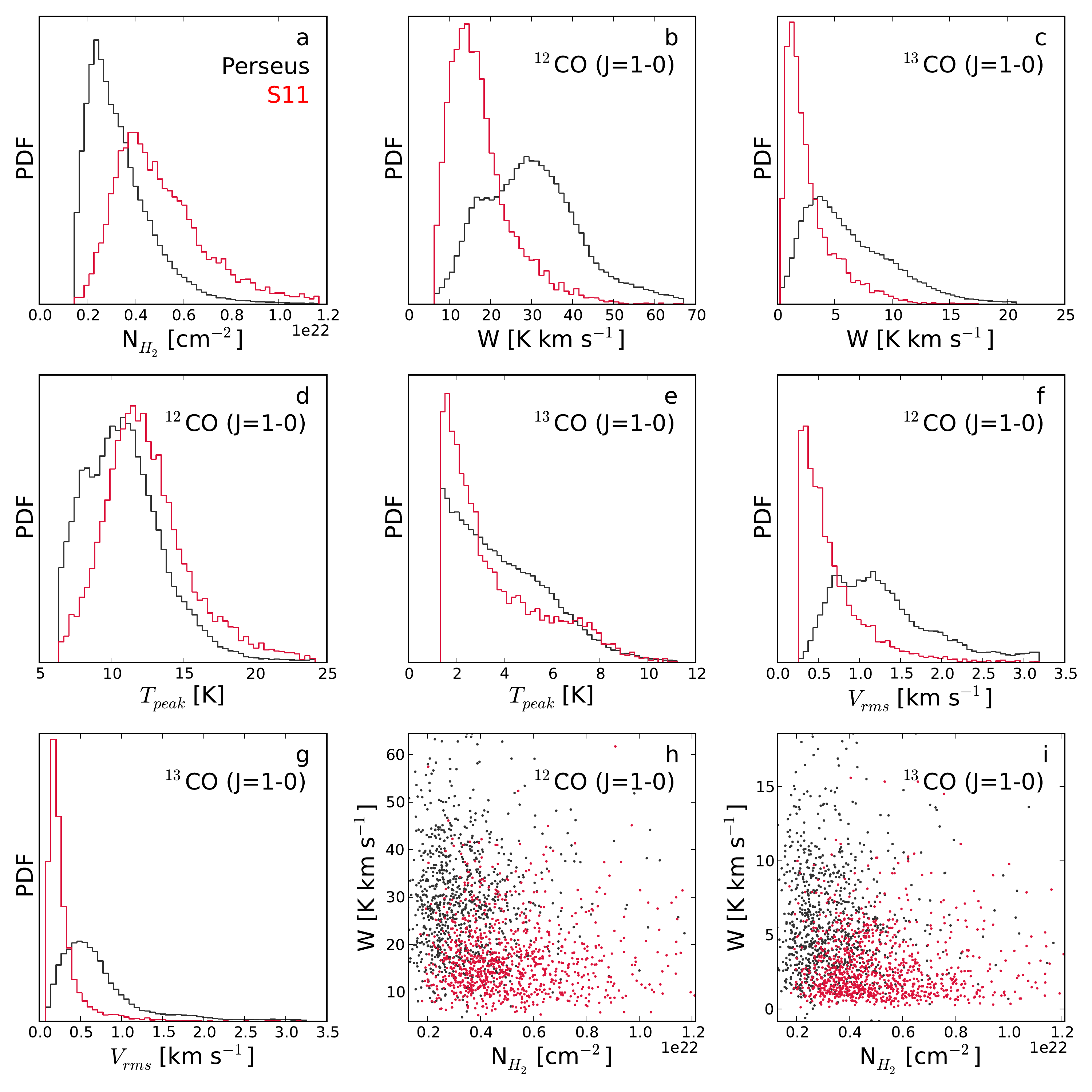}
\caption{The grid of diagnostic comparisons for the S11 simulation.}
\label{fig:dgrid_rahul}
\end{figure}

\section{Extracting cloud properties from Dendrograms}

The dendrogram algorithm defines a structure as a specific set of
connected voxels; we denote this set as $\Omega$. The intensity value at a given location $\vec{r}$ is denoted as $I(\vec{r})$ (this corresponds to the density for structures in a PPP cube). Here we describe how we measure properties from such a structure.

In all measurements, we define structures by contour surfaces and assume the structure ends at this boundary; that is, we adopt the ``bijection paradigm'' discussed in \cite{http://adsabs.harvard.edu/abs/2008ApJ...679.1338R}. This
assumption ``clips'' the low-intensity wings of structures embedded in ambient emission. 

\textbf{Location}

We compute the intensity-weighted first moment of each
structure to define its center.

\begin{equation}
\vec{\mu} = \frac{\sum_\Omega{I(\vec{r}) \cdot \vec{r}}}{\sum_\Omega{I(\vec{r})}}
\end{equation}  

\textbf{Orientation}

We compute the three moments of inertia of $I$; these vectors give the direction of
greatest and smallest elongation. We project the direction of greatest
elongation onto the PP plane, which defines the structure's major
axis $\hat{r}_{maj}$. The minor axis $\hat{r}_{min}$ is perpendicular to this.

\textbf{Size Scale}

We define the extent of each structure along the major and
minor axes to be the intensity-weighted second moment:

\begin{eqnarray}
\ell_{maj}^2 &=& \frac{\sum_{\Omega}{I(\vec{r}) \cdot \left[(\vec{r} - \vec{\mu}) \cdot \hat{r}_{maj}\right]^2}}{ \sum_\Omega{I(\vec{r})}} \\
\ell_{min}^2 &=& \frac{\sum_{\Omega}{I (\vec{r}) \cdot \left[(\vec{r} - \vec{\mu}) \cdot \hat{r}_{min}\right]^2}}{ \sum_\Omega{I(\vec{r})}} \\
\ell_r &=& \sqrt{\ell_{maj} \times \ell_{min}} \\
A &=& \ell_r^2
\end{eqnarray}

Another common method for measuring the size scale of irregular PPV structures is to measure the area by counting the
number of distinct (X, Y) pixels that a structure occupies, and defining $\ell_r' = \sqrt{A/\pi}$. $\ell_r'$ tends to be about $50\%$
larger than $\ell_r$, since the latter measure is intensity-weighted, and structures are usually centrally-condensed.

When measuring the virial parameter, we multiply $\ell_r$ by 1.91 to correct for this central concentration. This is the same factor
applied and discussed in \cite{http://adsabs.harvard.edu/abs/2008ApJ...679.1338R}.

\textbf{Velocity Dispersion}

We define the velocity dispersion $v_{\rm rms}$ as the second moment of the intensity distribution along the velocity direction.

\textbf{Mean Intensity}
  
  We simply compute the mean of the intensity for all voxels belonging to a structure $\Omega$.
  
 \textbf{Mass}
 
For PPP structures, mass can be computed directly by integrating the density field. For PPV structures, we assume that a structure's mass is linearly proportional 
to its integrated intensity (that is, we adopt the ``X-factor'' assumption; \citealt{http://adsabs.harvard.edu/abs/2008ApJ...679..481P}).
Such a proportionality exists if emission is optically thin and the emitting molecule has a constant abundance and excitation temperature. None of these assumptions holds for the simulations in this paper, and the X-factor varies as a function of position. Furthermore, since these simulations are under-luminous compared to real clouds, it would be unwise to use standard X-factors quoted in the literature. Instead, we set the conversion factor individually for each simulation, to best recover the input mass from synthetically-observed bright emission. We look at the brightest 5\% of the lines-of sight, and define $X^{\rm syn}$ as the mean of the ratio of surface density / integrated CO intensity in these pixels. We then estimate mass as
\begin{equation}
M_{\rm PPV} = \sum_{\Omega} I(\vec{r}) \delta v \delta x^2 X^{\rm syn}
\label{eq:xfactor_mass}
\end{equation}
where $\delta v$ is the velocity width of a pixel, and $\delta x$ is the length of a pixel.

The brightest 5\% of pixels represent the densest regions of each simulation, where the opacity is highest. The X factor derived from these pixels tends to over-estimate masses from less-opaque but equally-excited lines of sight, with lower mass-to-light ratios. Likewise, it underestimates the mass for the faintest lines of sight, where CO is sub-thermally excited and the mass-to-light ratio is large.
\end{document}

%% file: terms_table.tex
\begin{deluxetable*}{ll}
\tablecolumns{3}
\tablewidth{0in}
\tabletypesize{\scriptsize}
\tablecaption{Terminology}
\tablehead{\colhead{Term} & \colhead{Description}}
\startdata
Density or PPP structure/feature & A contiguous volume in real, PPP space. Defined by a 3D density contour.\\
Intensity or PPV structure/feature & A contiguous region in PPV space. Defined by intensity contours in a spectral line observation. \\
Density field & The density at each PPP location in a simulation \\
Velocity field & The (line-of-sight) velocity at each PPP location in a simulation \\
Intensity field & The intensity of a spectral line at each PPV location in a simulation \\
Confusion & General term for the imperfect correspondence between PPP structures and PPV structures. \\
\enddata
\label{tab:terms}
\end{deluxetable*}

%% file: sim_params_table.tex
\begin{deluxetable*}{lll}
\tablecolumns{3}
\tablewidth{0in}
\tabletypesize{\scriptsize}
\tablecaption{Summary of each simulation}
\tablehead{\colhead{} & \colhead{S11} & \colhead{O1}}
\startdata
Box Size & 20 pc & 25 pc \\
Simulation Code & Zeus-MP & ORION \\
Gridding & 256$^3$ & 256$^3$ + 4 levels of AMR refinement \\
Driven Turbulence & Yes & Yes \\
Driving Power Spectrum & Uniform $1 < k < 2$ & Uniform $1 < k < 2$ \\
Gravity & No & Yes \\
B field & 5.85 $\mu$G & 0 \\
Gas Temperature & Variable (10-200K) & 15K \\
Chemistry & H, O, C & None \\
Background UV & 2.7e-3 erg cm$^{-2}$ s$^{-1}$ & No \\
Constant CO/H$_2$ Abundance & No & 1.75 e-4 \\
$^{12}$CO/$^{13}$CO abundance & 70 & 70 \\
Radiative Transfer Code & RADMC 3D & RADMC 3D \\
Microturbulence & 0.2 km s$^{-1}$ & 0.2 km s$^{-1}$ \\
Metallicity & Solar & N/A \\
Mean number density ($n_{\rm H}$) & 100 cm$^{-3}$ & 58 cm$^{-3}$ \\
Mach Number & $\sim 6$ & 22 \\
Isothermal & No & Yes \\
Output time(s) & 5.7 Myr & 2.5 Myr (with gravity)\\
Mass in stars & N/A & 722 M$_{\odot}$ (2.4\%)
\enddata
\label{tab:sim_params}
\end{deluxetable*}

%% file: diagnostic_table.tex
\begin{deluxetable}{lrr}
\tablecolumns{3}
\tablewidth{0in}
\tabletypesize{\scriptsize}
\tablecaption{Diagnostic scores for the O1 and S11 simulations}
\tablehead{\colhead{Category} & \colhead{O1 Score} & \colhead{S11 Score}}
\startdata
Column Density & \textbf{0.87} & 0.52 \\
W($^{12}$CO 1-0) & 0.36 & \textbf{0.47} \\
W($^{13}$CO 1-0) & \textbf{0.68} & 0.53 \\
$^{12}$CO 1-0 Velocity Dispersion & \textbf{0.84} & 0.47 \\
$^{13}$CO 1-0 Velocity Dispersion & \textbf{0.91} & 0.41 \\
Peak $^{12}$CO 1-0 intensity & 0.68 & \textbf{0.77} \\
Peak $^{13}$CO 1-0 intensity & 0.75 & \textbf{0.84} \\
N$_{col}$ vs W($^{12}$CO 1-0) & \textbf{0.60} & 0.50 \\
N$_{col}$ vs W($^{13}$CO 1-0) & \textbf{0.78} & 0.53 \\\enddata
\label{tab:scores}
\end{deluxetable}